# Analysis of Stakeholder Involvement in Nuclear Power Plant Cost Overruns and Implications for Contract Structuring

August 1, 2026

Christopher Forsyth[1,*], Levi M. Larsen[2], Ryan Spangler[3], Chandu Bolisetti[3], Jason Hansen[3], Botros Hanna[3], Abdalla Abou-Jaoude[3], Jia Zhou[4], and Koroush Shirvan[1]

[1]Massachusetts Institute of Technology, Cambridge, MA, USA
[2]Norwegian University of Science and Technology, Trondheim, Norway
[3]Idaho National Laboratory, Idaho Falls, ID, USA
[4]Argonne National Laboratory, Argonne, IL, USA
*Corresponding Author: chris24@mit.edu

## Abstract

Existing evidence on nuclear power plant construction suggests that individual stakeholders involved in project delivery are often not responsible for all of the cost overruns in their scope of the project, which can lead to costly and time-consuming litigation to determine who is truly at fault for overruns. This study introduces a framework to model the share of overruns caused and received as payment by five stakeholders in a nuclear construction project: equipment suppliers, construction subcontractors, the design and management team, creditors, and the nuclear regulator. We then perform a contract analysis under three common contract structures – fixed-price, cost-plus, and performance-based – to show the profit misallocations that can result, revealing the advantages and disadvantages of each contract type for aligning stakeholder incentives. Regardless of the contract type chosen, strong owner involvement is crucial for project success, and we conclude with specific recommendations for project owners seeking to minimize cost overruns.

## Introduction

Recent nuclear construction projects, particularly in the Western world (United States, France, Finland, and United Kingdom), have seen major cost and schedule overruns – often more than twice the original estimates [1] – that have dampened investor interest and prevented nuclear power from gaining a larger share of the energy market. Since construction costs comprise most of the levelized costs of nuclear energy, this work aims to provide a better understanding of the origins of nuclear construction cost overruns so they can be more effectively mitigated through effective contracts in the future. This article specifically examines how five stakeholders involved in nuclear construction – construction subcontractors, equipment suppliers, design and management teams, creditors, and the nuclear regulator – play a role in cost and schedule overruns.

Existing evidence on megaproject overruns [2] [3] [4] [5] supports the idea that individual stakeholders involved in the project are not always responsible for all of the cost overruns they charge to the project owner. The mistakes or shortcomings of one stakeholder can affect the performance of downstream stakeholders, sometimes resulting in lost profits. This is often met with costly and time-consuming litigation between parties to determine who is at fault and recover said profits.

Current academic literature has thoroughly discussed both the prevalence and magnitude of cost overruns in nuclear construction, and some work also attempts to identify the causers of these overruns. However, less is said about how blame for overruns is ultimately attributed, or misattributed, on a cost basis. This paper aims to address this gap in the literature by illustrating how the mismatches between overruns caused and overruns received as payment by each stakeholder can affect their profit outcomes under three types of contract frameworks: performance-based, fixed-price, and cost-plus. By modeling stakeholder involvement in overruns and performing this contract analysis exercise, we aim to generate useful takeaways for nuclear power plant developers, owners, and construction managers about how to structure contracts and execute their projects effectively.

Thus, our analysis examines nuclear construction cost overruns through two different lenses: (1) who causes the overruns, and (2) who charges the cost of overruns to the project owner. Figure 1 helps to illustrate the distinction between these two concepts. In the “project execution diagram,” which shows the flow of goods and services necessary for project completion, stakeholders are connected to each other in different ways than in the “cash flow diagram,” which shows the flow of money from the owner to other stakeholders. This can cause delays and cost escalation in the work scope of stakeholders whose work depends on the performance of upstream parties.

For example, suppose an equipment supplier discovers an error in the tolerances on a component drawing provided by the reactor vendor during fabrication, necessitating the return of the drawing to the vendor for revisions. Figure 1 shows how this mistake ripples through the project ecosystem. In the project execution diagram, the disruption flows from the design and management teams to the equipment suppliers, which affects the productivity of the construction subcontractors, and by extension, slows the completion of the power plant. The cash flow diagram shows how compensation for these impacts flows in the opposite direction, from the owner to the stakeholders involved in power plant completion. The owner must pay the construction subcontractors for additional hours spent waiting for component delivery, the equipment suppliers for factory time spent unproductively, and the design and management team for engineering hours spent on revising the drawings. The owner will also pay increased financing charges to the creditors (banks, equity financers, etc.) while the plant completion is delayed. Thus, even though the reactor vendor was solely responsible for causing this cost overrun, many other stakeholders are affected and must receive payment for overruns.

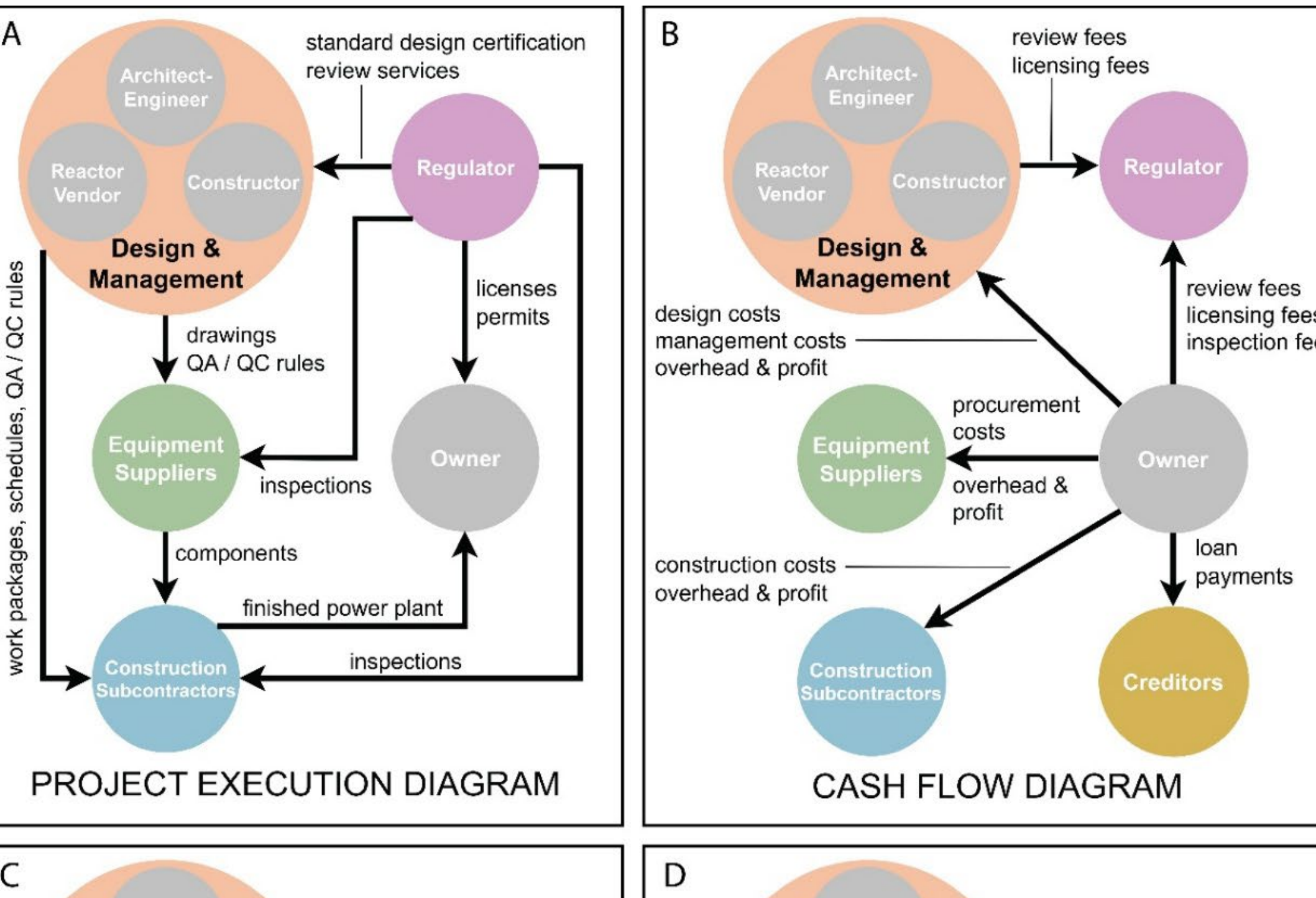

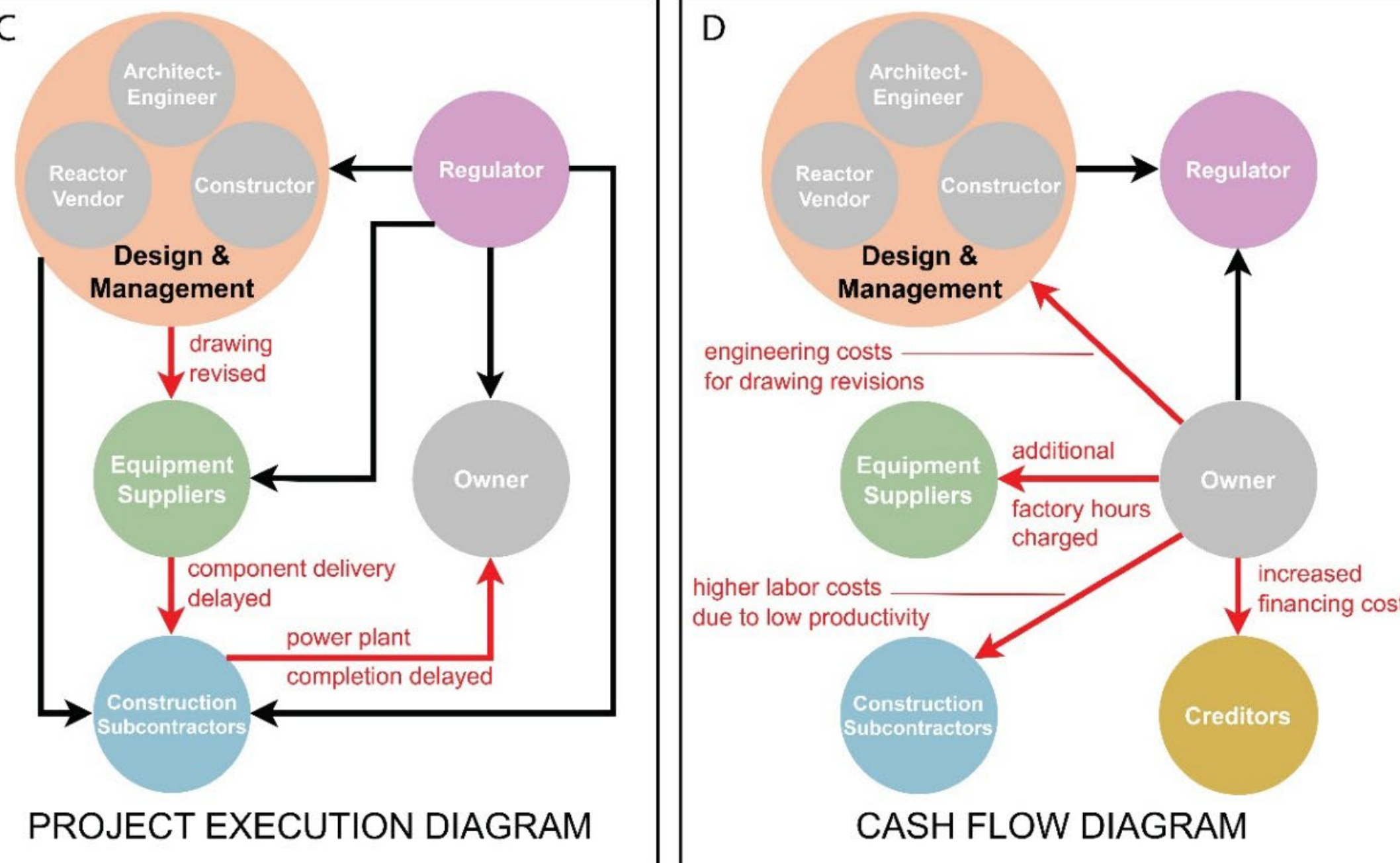

**Figure 1. Connections between stakeholders in a nuclear construction project.**
(A) Project execution diagram showing the flow of goods and services necessary for project completion; (B) cash flow diagram showing the flow of payments from the owner to the various stakeholders involved in plant design and construction; (C) red arrows show how an error in a drawing issued by the reactor vendor trickles down to affect the work scope of other stakeholders; (D) red arrows show how each stakeholder is compensated for this disruption. The setup in this illustrative case is based on U.S. nuclear construction experience, but other arrangements between stakeholders are possible, especially in vertically integrated nuclear industries where multiple functions (design, procurement, construction management, etc.) are performed by a single entity.

## Literature Review

Many previous studies have attempted to quantify the sources of nuclear power plant cost escalation and cost overruns, although most have not approached this question from the perspective of stakeholder attributions or contracting theory with quantitative modeling. A study by Eash-Gates et al. [7] cited historic reports by Energy Economic Data Base (EEDB) [8] that showed a large portion (72%) of nuclear construction cost escalation in the United States over the period from 1976-87 was attributable to a substantial rise in indirect costs, including "home office engineering services (engineering design, purchasing and expediting, cost control, and planning and scheduling), field job supervision (salaries and relocation expenses), temporary construction facilities (materials and labor to construct and manage buildings needed during construction), and payroll insurance and taxes." Most of the capital cost increases over this period were caused by "those activities and practices related to meeting accountability type requirements," including "regulatory reviews, design review, project control, analysis verification, procedure development and implementation, equipment qualification, inspection, testing and similar or related activities."

Eash-Gates et al. also found that labor productivity in U.S. nuclear construction projects has continuously declined, both during the original construction boom of the 1970s-80s and in recent projects. It is notable that the observed productivity declines are worse than those observed in the U.S. construction sector as a whole [6] [7], indicating that the nuclear industry has been subject to uniquely severe productivity disruptions over its history, such as the string of design changes mandated by the Nuclear Regulatory Commission (NRC) in the middle of plant construction during the 1970s-80s [9] and the low levels of design completion that have characterized recent Western projects [6]. Thanks to the massive rise in indirect costs, rework, and low labor productivity, a larger and larger share of Western nuclear power construction costs is attributable to labor rather than materials or equipment costs: the final EEDB update found that the cost of a large pressurized water reactor with median U.S. construction experience ("PWR12-ME") in 1987 was roughly two-thirds labor and one-third materials costs, and the labor share has presumably become even larger in the 21st century as U.S. nuclear construction productivity has continued to decrease [7].

Studies examining which specific stakeholders are responsible for which shares of cost overruns are more limited. Previous econometric analyses have examined the influence of a few high-level stakeholders on plant costs, particularly the reactor vendor that provides the nuclear steam supply system, the architect-engineer (A-E) that designs the balance of plant, the constructor (sometimes called the "prime contractor") that manages construction activities, and the owner (e.g., a utility company) that orders the plant. The Energy Information Administration (EIA) found a 35% reduction in specific overnight capital costs (OCC) for projects in which the owner acted as its own constructor rather than hiring an external construction management firm [10]. Jansma and Borcherding likewise found lower costs for projects where the owner acted as its own constructor, but also found that costs were higher when the owner acted as its own A-E [11]. Komanoff found a decline of 7% in both specific overnight capital costs (OCC) and construction duration for each doubling in the number of plants constructed by a given A-E firm [9]. Komanoff also found that for the six reactors in the 46-reactor cost sample for which the utility designed and built the plant itself, costs were higher than for other plants, though this study was published earlier than the EIA study [10] and the sample size of completed reactors was therefore smaller than EIA's (67 reactors, of which 18 had the utility acting as its own constructor). The EIA also found that constructors exhibit a positive learning effect (i.e., cost reductions due to experience building multiple plants), but the effect was only statistically significant for plants in which the owner acted as its own constructor, suggesting that learning-related cost savings may not be passed on to the owner if they hire an external constructor. These econometric studies shed useful light on the learning curves of some stakeholders and the impact of different project management regimes on costs. However, they do not include several additional stakeholders that we consider in our analysis (most notably equipment suppliers, construction subcontractors, and creditors – see Figure 1), nor do they attempt to quantify the specific cost overrun shares attributable to each individual stakeholder.

Many authors in the project management discipline have studied cost data and qualitative evidence from specific projects to determine how different stakeholders are involved in overruns. Locatelli et al. in particular have conducted thorough analyses and literature reviews of stakeholder involvement in nuclear power plant construction cost overruns [12] [13]. They find that nuclear power plants represent a highly complex construction megaproject requiring the coordination of a uniquely large number of stakeholders, including government bodies, contractors, and local communities. These stakeholders can cause cost overruns through a variety of avenues, including poor organizational structuring or incentive frameworks; poor contract management and improper risk allocation between stakeholders; and failure to engage proactively and productively with the community, political decision-makers, and regulators. The authors recommend a number of approaches to avoid these megaproject pitfalls, including early and continuous engagement with external stakeholders, better "front-end loading" (i.e., more thorough planning before construction start), more carefully structured contracts with a fair allocation of risk, and the use of "special purpose entities" (SPEs) to deliver specialized construction megaprojects like nuclear power plants, among others.

As our examination of the literature highlights, other authors have examined the general question of where nuclear construction cost overruns tend to come from using econometric methods and case studies, and have found that poor project management and a misalignment of incentives between stakeholders are important drivers of cost overruns. In this paper, we offer a theory that may help to explain one of the root causes of these misaligned incentives – namely, that the stakeholders who cause overruns are not the same as those who receive payment for overruns, and this results in some stakeholders being penalized for the shortcomings of others.

## Analytical approach

We pursue a new approach to studying stakeholder involvement in nuclear cost overruns that does not rely on econometric regression or qualitative case studies of past projects. Instead, we begin with a pair of empirically derived bottom-up cost models from United Engineers & Constructors' Energy Economic Data Base (EEDB) [8] for an 1144-MWe pressurized water reactor entering service in the U.S. in 1987: the PWR12-ME (Median Experience), which reflects the median construction experience in the U.S. at that time, and the PWR12-BE (Better Experience), which reflects costs for "a small group of single unit nuclear power plants at the low end of the range of base construction costs for single unit nuclear power plants currently entering service" in 1987 [8]. The PWR12-BE has an overnight capital cost of $1.456 billion (1987 USD) and a construction duration of 72 months, while the PWR12-ME has an overnight capital cost of $2.529 billion (1987 USD) and a construction duration of 98 months. We then establish a series of mechanistic, historically-informed equations that allow us to create a stylized scenario in which we model:

(1) How the overruns between the PWR12-BE and ME arose mechanistically;
(2) Which stakeholders caused the overruns;
(3) Which stakeholders received payment for the overruns;
(4) How mismatches between overruns caused and overruns received by each stakeholder affect the stakeholders' incentive structures and profit margins under different contract frameworks.

For the first step, we borrow a set of equations from the Nuclear Reactor Capital Cost Reduction Tool (henceforth referred to as the Cost Reduction Tool or CRT) developed by Idaho National Laboratory (INL) in 2024 [14], which introduced simple formulations for modeling cost overrun mechanisms to determine how nuclear construction costs might be reduced across multiple reactor deployments under different circumstances. The CRT has been used to model the influence of several factors on nuclear power plant construction costs, including stakeholder proficiency levels, reactor design maturity, the level of cross-site standardization across multiple projects, and the level of detailed design completion prior to construction start, among others. For our analysis, we focus on three specific mechanisms for OCC and schedule overruns modeled by the Cost Reduction Tool – rework, low construction site productivity, and supply chain delays – and we also capture financing cost overruns driven by these three sources of OCC and schedule overruns. It should be noted that throughout this analysis, we use the PWR12-BE as our cost baseline, and the term "overruns" is used to describe any excess costs and construction duration exceeding PWR12-BE cost and schedule outcomes.

For the second and third steps, we propose a new framework to model the shares of overruns caused by individual stakeholders and the shares received by each stakeholder as payment. This new stakeholder attribution framework is partially informed by the underlying structure of the Cost Reduction Tool equations and the EEDB reactor cost models. For example, since direct costs in EEDB are subdivided into site material costs, site labor costs, and factory equipment costs, we assume all site material and site labor costs are received by the Construction Subcontractors stakeholder as payment, while all factory equipment costs are received by the Equipment Suppliers stakeholder (see *Methods* § *Overruns by recipient*). The attribution framework is also partially informed by additional empirical data from U.S. nuclear construction projects. For example, we model which stakeholders are responsible for low productivity on the construction worksite using data from a 1980 survey of craft laborers across 11 U.S. nuclear construction projects [15] (see *Methods* § *Overruns by causer*). The stakeholder attribution exercise enables us to illustrate the mismatch that can arise between overruns caused and overruns received as payment by individual stakeholders in nuclear construction projects.

For the fourth step, we examine the implications of this mismatch under three contract frameworks used commonly in construction megaprojects: performance-based contracting, fixed-price contracting, and cost-plus contracting. The results illustrate the sharp discrepancies in profit outcomes that can occur due to mismatches between overruns caused and overruns received, potentially resulting in disputes and incentive misalignment among stakeholders. We conclude with takeaways about the advantages and disadvantages of each contract type for multi-stakeholder projects, as well as providing recommendations for minimizing cost overruns and maximizing incentive alignment between stakeholders under each contract type.

# Results

This section presents the results of the stylized scenario we constructed to analyze stakeholder involvement in nuclear construction cost and schedule overruns. As discussed in *Introduction* and *Methods*, our model analyzes the overruns in cost and schedule between the PWR12-BE reactor cost model for a well-executed nuclear construction project and the PWR12-ME reactor cost model for a poorly-executed nuclear construction project [8]. In the *Stakeholder analysis* section, we provide breakdowns of the overruns between the PWR12-BE and the PWR12-ME by mechanism (rework, low productivity, supply chain delays, and financing overruns) and then highlight which of the five stakeholders considered in our model (see Figure 1) were responsible for causing the overruns and which stakeholders received the overrun cash flows as payment. In the *Contract exercise* section, we show how mismatches between overruns caused and overruns received by each stakeholder can affect their profit outcomes under three types of contract frameworks: performance-based, fixed-price, and cost-plus.

## Input assumptions

Table 1 summarizes the key input assumptions for the CRT equations used in the stakeholder analysis. The productivity, $p$, represents the assumed difference in labor productivity between the PWR12-BE and the PWR12-ME, which is used to calculate the amount of overruns caused by low labor productivity. As described in *Methods*, we assume the labor productivity of the PWR12-ME is 75% of the PWR12-BE productivity, an assumption informed by difficult construction experience at Vogtle 3 & 4, in which the "Cost Performance Index" ($CPI$ = hours spent / hours earned) was reported to be ~1.3 - 1.4 as of 2020-21 [16] [17], corresponding to a labor productivity of 1 / $\mathrm{CPI} \approx$ 0.75.

**Table 1. Key input assumptions used for the stakeholder analysis.**

| Input | Notation | Input Value | Relevant Equation(s) |
|---|---|---|---|
| productivity | $p$ | 0.75 | 1, 5 |
| architect-engineer proficiency lever (0 – 2) | $AEP$ | 1.0 | 3 |
| construction service proficiency lever (0 – 2) | $CSP$ | 1.5 | 4 |
| architect-engineer rework factor | $R_{AE}$ | 1.127 | 2 |
| construction service rework factor | $R_C$ | 1.063 | 2 |
| design incompletion rework factor [a] | $R_{design}$ | 1.107 | 2 |
| financing rate | $r$ | 4% | S3, S4, S5 |

[a] Back-calculated from the total rework factor ($R_{tot}$ = 1.327) as described in *Methods* and *Supplemental Information*.

The architect-engineer proficiency lever ($AEP$) and the construction service proficiency lever ($CSP$) are CRT inputs that can be adjusted to reflect the assumed competency of the A-E (a lower-level stakeholder that is part of the Design & Management team; see Figure 1) and the Construction Subcontractors, respectively. These levers, which may take on a value between 0 (minimum proficiency) and 2 (maximum proficiency), are used to calculate "rework factors" which are then used to model the degree to which the Design & Management, Construction Subcontractors, and Regulator stakeholders are responsible for rework-related overruns (see *Methods* § *Overrun mechanisms* § *Rework* for more details). For our illustrative scenario, we choose values of $AEP$ = 1.0 (medium A-E proficiency) and $CSP$ = 1.5 (medium-high construction service proficiency), resulting in rework factors of 1.127 and 1.063, respectively. The design incompletion rework factor, which is assumed to be caused by the Regulator (see *Methods*), is back-calculated from $R_C$ and $R_{AE}$ as described in the *Overrun mechanisms* § *Rework* section to be 1.107. With these three rework factor values, the resulting shares of responsibility for rework OCC overruns are 42.7% for Design & Management, 21.2% for the Construction Subcontractors, and 36.1% for the Regulator.

Lastly, we assume a financing rate, $r$, of 4% for calculating the cost of interest during construction (IDC), which allows us to model additional financing cost overruns that result from the OCC and schedule overruns between the PWR12-BE and PWR12-ME. A cost of capital of 4% is very low compared to the actual interest rates and required returns on equity that would have applied in the 1980s (the decade in which the PWR12-BE and PWR12-ME reactor cost models are based), but may be more representative of the cost of capital for more modern U.S. nuclear projects where interest rates are lower and inexpensive government financing is available. We choose to model a financing rate of 4% for the sake of this stylized scenario, but choosing a higher financing rate will generally result in the Creditors stakeholder receiving a larger share of the cost overruns in Figure 3(C).

## Stakeholder analysis

First, we present the modeled overruns by mechanism. Figure 2(A) shows that under the assumptions chosen for our analysis (*Results* § *Input assumptions*), 30.1% of the $1.31 billion (1987 USD) in cost overruns between PWR12-BE and PWR12-ME is caused by low worksite productivity, 47.7% is caused by rework, and 4.0% is caused by supply chain delays, while 18.2% is caused by financing overruns tied to these three sources of OCC and schedule escalation, as reflected in Figure 3(A). As for construction duration, 32.2% of the 26-month schedule overrun is due to low productivity, 31.7% due to rework, and 36.1% due to supply chain delays, as shown in Figure 4(A).

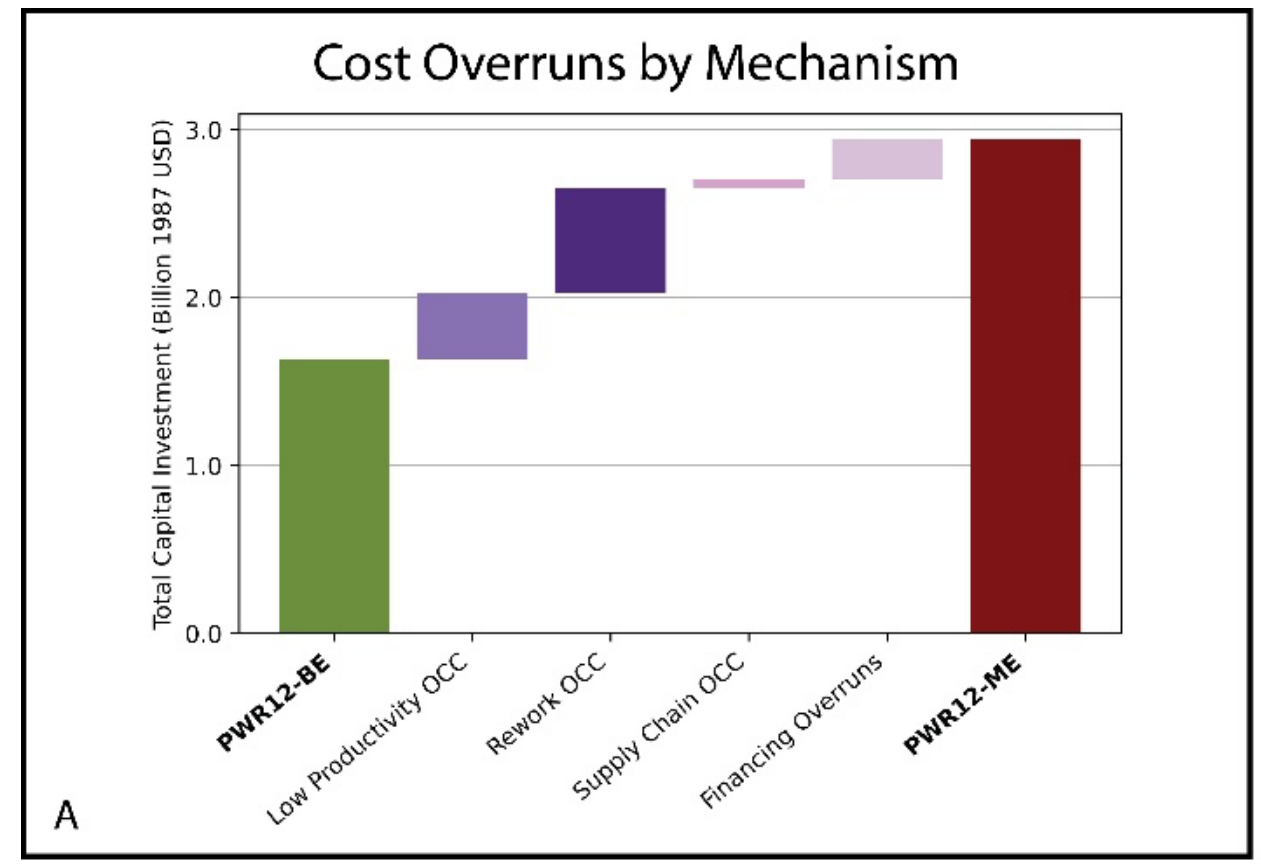

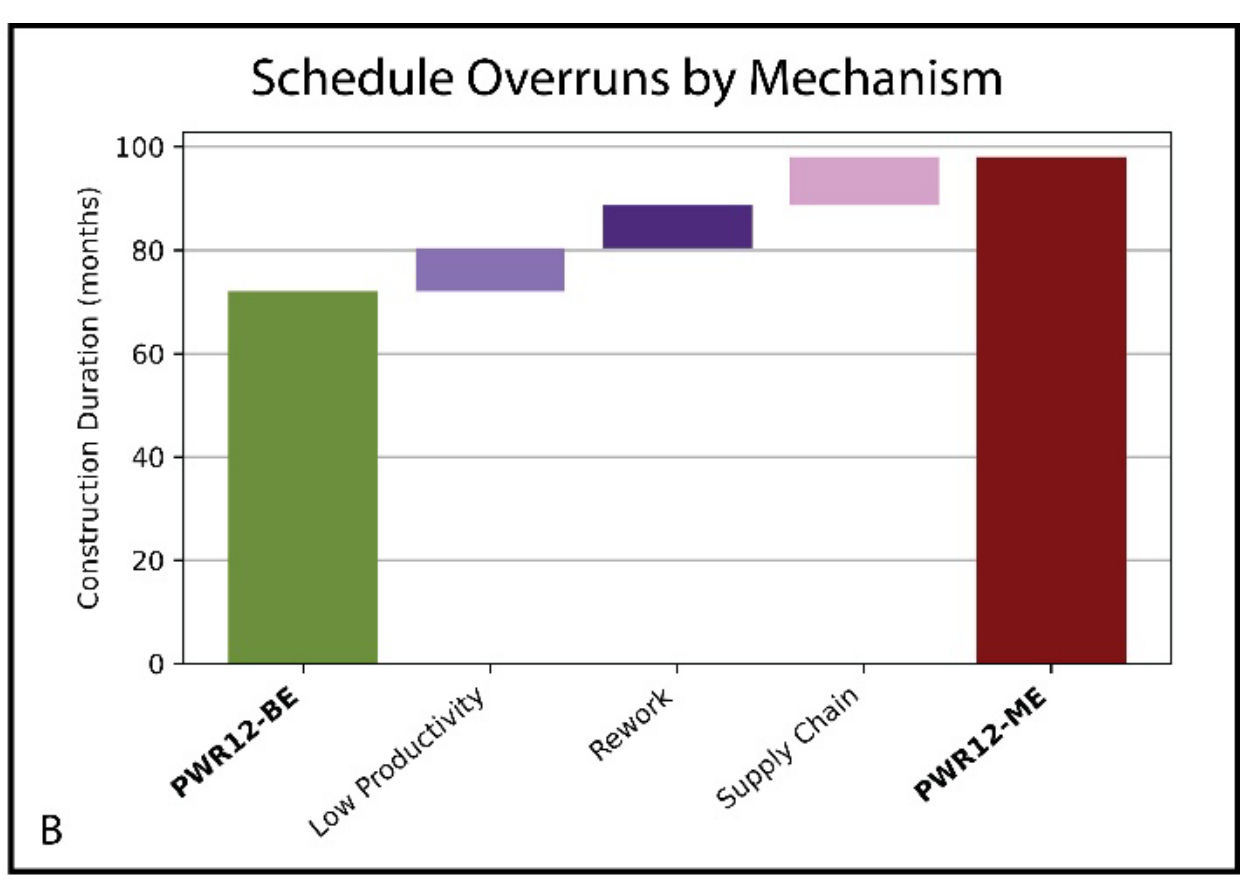

**Figure 2. Modeled sources of cost and schedule overrun between the PWR12-BE and PWR12-ME.** (A) Sources of cost overrun, including OCC overruns due to low worksite productivity, rework, and supply chain delays, as well as financing overruns driven by OCC and schedule overruns; (B) sources of schedule overrun, including low worksite productivity, rework, and supply chain delays.

Having modeled the mechanisms by which overruns occur, we next move to the results of the stakeholder analysis, in which we assess which of the five stakeholders we considered in our model are responsible for causing these overruns, and which stakeholders receive payments for the cost overruns. Figures 3(B)-(C) and 4(B) show the outcome of this analysis. In Figure 3(B), the $1.31 billion (1987 USD) in cost overruns is shown to be caused by a combination of Design & Management (40.6%), Construction Subcontractors (26.1%), Regulator (22.2%), and Equipment Suppliers (11.2%), while the Creditors are not responsible for causing any cost overruns because they are not directly involved with the execution of the project (other than to lend money). Meanwhile, Figure 3(C) shows that the Creditors collect a share of the cost

overruns as payment (18.2%). This share is relatively low due to the assumed financing rate of 4.0%. The Regulator does not receive any portion of the cost overrun as payment because we assume any regulatory fees are negligible compared to the multi-billion dollar cost of the overall project. Finally, Figure 4(B) shows how much of the 26 months in schedule overruns between the PWR12-BE and PWR12-ME are caused by each stakeholder: 40.1% by the Equipment Suppliers, 27.8% by Design & Management, 19.1% by the Construction Subcontractors, and 13.0% by the Regulator.

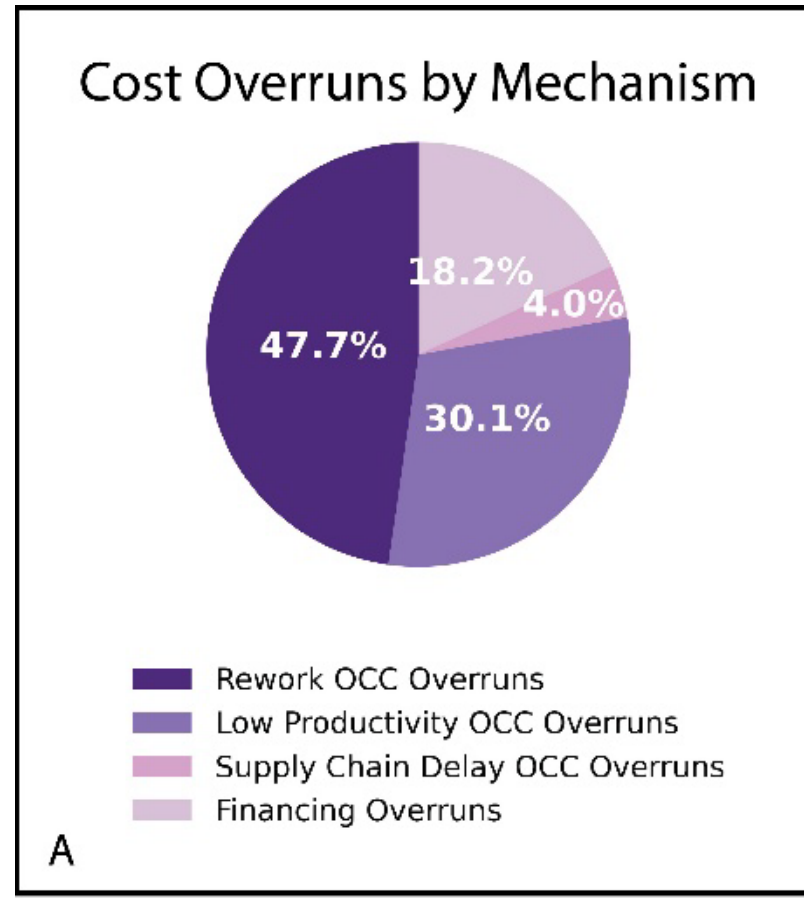

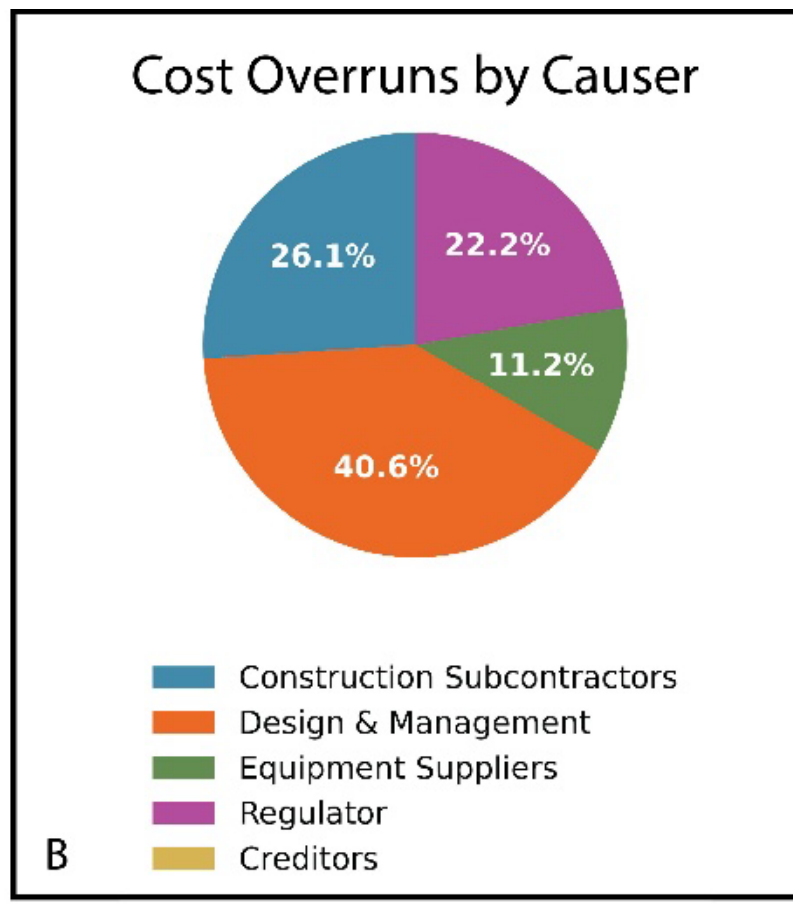

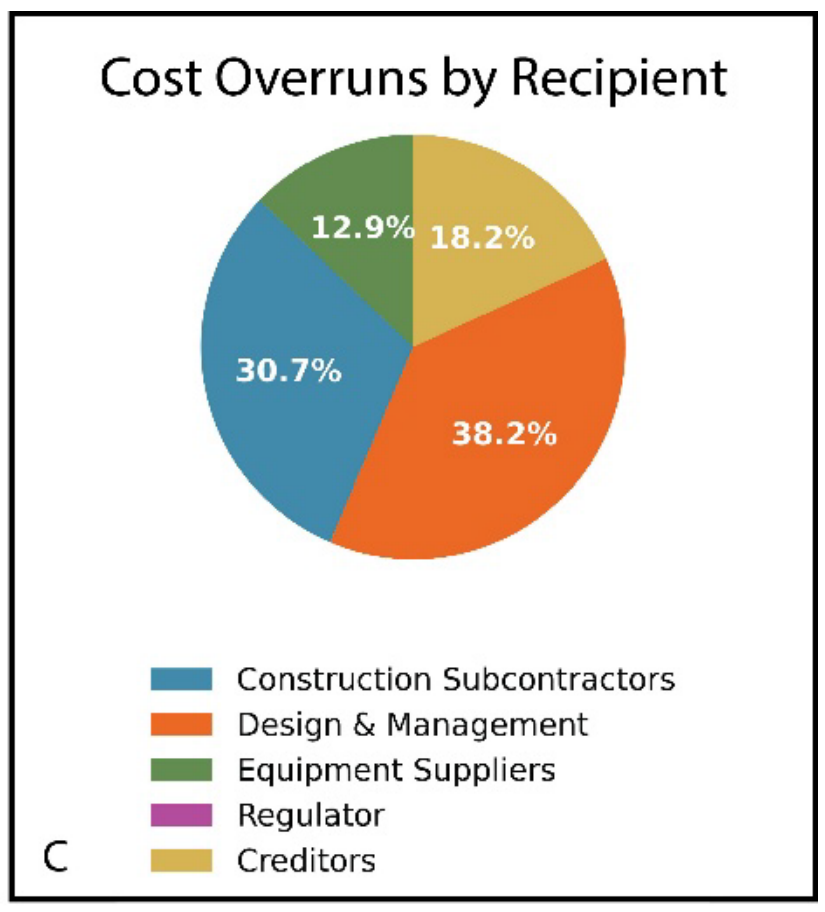

**Figure 3. Breakdown of cost overrun results from the stakeholder analysis.**
(A) Cost overruns between PWR12-BE and PWR12-ME broken down by mechanism (rework OCC, low productivity OCC, supply chain delay OCC, and financing); (B) breakdown of which stakeholders caused the cost overruns; (C) breakdown of which stakeholders received payment for the cost overruns.

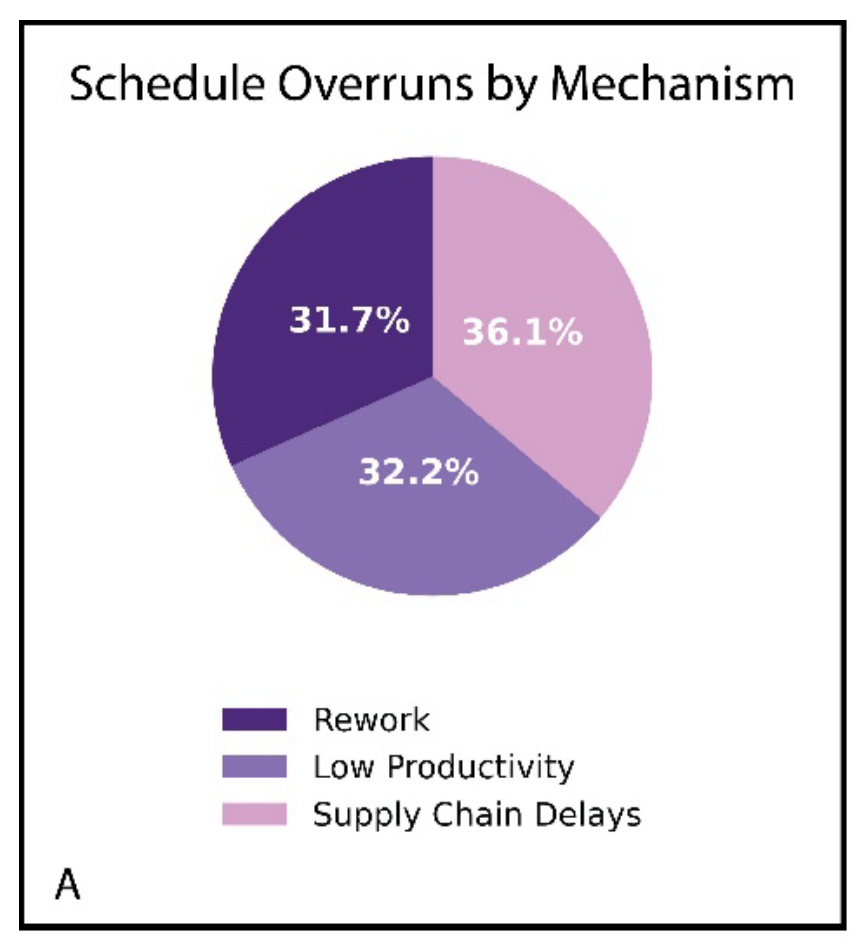

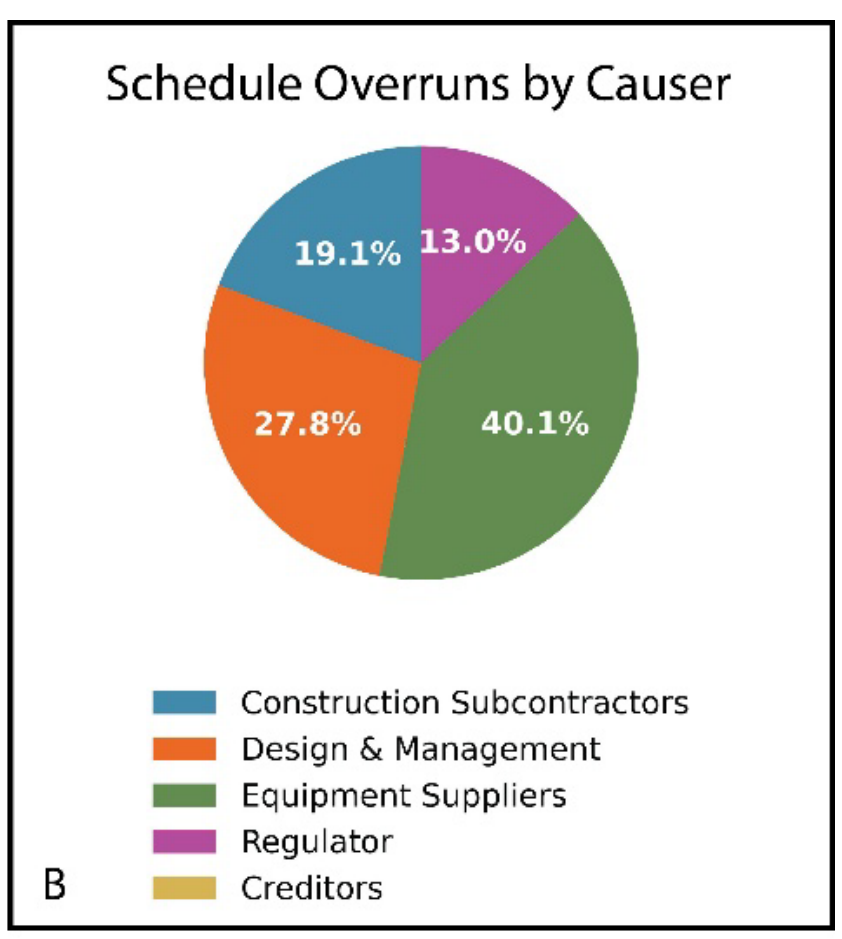

**Figure 4. Breakdown of schedule overrun results from the stakeholder analysis.**
(A) Schedule overruns between PWR12-BE and PWR12-ME broken down by mechanism (rework, low productivity, and supply chain delays); (B) breakdown of which stakeholders caused the schedule overruns.

Of the three stakeholders who both caused and received payment for overruns (Design & Management, Construction Subcontractors, and Equipment Suppliers), two stakeholders received payment for more overruns than they caused (Construction Subcontractors and Equipment Suppliers), while Design & Management caused more overruns than they received payment for. It makes intuitive sense that Design & Management is responsible for causing an outsized share of the cost overruns since this stakeholder is upstream from many of the other stakeholders (see Figure 1), and can therefore easily cause overruns in other stakeholders' work scopes. It is worth noting that these results are sensitive to the assumptions in Table 1. For instance, if the construction service proficiency ($CSP$) is lowered from 1.5 (medium-high proficiency) to 1.0 (medium proficiency), then the Construction Subcontractors cause 38.2% of the cost overruns while only collecting payment for 30.7% of the overruns.

The share of schedule overruns caused by the Equipment Suppliers (40.1%) is large compared to the share of cost overruns caused by this stakeholder (11.2%). This is because although late deliveries by the Equipment Suppliers impact financing costs due to a longer project duration over which interest can accrue, financing cost overruns only make up 18.2% of the total cost overruns due to the low financing rate of 4.0%, and the Equipment Suppliers only cause 19.3% of the total financing overruns because their impact on OCC is so low relative to other stakeholders. Thus, the impact of late equipment deliveries on financing costs only accounts for 18.2% × 19.3% = 3.5% of the overall cost overruns. Supply chain delays also have an impact on OCC by increasing the construction duration in Equations S9 and S10, and this effect accounts for 4.0% of the total cost overruns as shown in Figure 3(A). Lastly, late deliveries by the Equipment Suppliers are modeled to be responsible for 12.35% of the low worksite productivity (see Figure 6), which makes up 30.1% of total cost overruns, so the Equipment Suppliers cause an additional 12.35% × 30.1% = 3.7% of the total cost overruns through this mechanism. Thus, the Equipment Suppliers cause a total of 3.5% + 4.0% + 3.7% = 11.2% of the overall cost overruns.

It is also interesting to note that the presence of one stakeholder who causes but does not receive overruns (the Regulator) and another stakeholder who receives but never causes overruns (the Creditors) virtually guarantees a mismatch between overruns caused and overruns received for the other three stakeholders. Since the share of overruns caused by the Regulator in Figure 3(B) is 22.2% and the share of overruns received as payment by the Creditors in Figure 3(C) is 18.2%, the remaining three stakeholders must collectively be responsible for causing 77.8% of the overruns and must receive payment for 81.8% of the overruns. This makes it mathematically impossible for the Equipment Suppliers, Construction Subcontractors, and Design & Management to each be responsible for causing as much of the overruns as they receive payment for, and this will always be true (unless the Creditor and Regulator shares coincidentally happen to match). This potentially suggests that mismatches between overruns caused and received as payment by individual stakeholders in nuclear construction projects are not only possible, they are likely, which may help to explain why disputes and litigation over responsibility for cost overruns occur so often in these megaprojects, including in the recent U.S. builds at Vogtle 3 & 4 and V.C. Summer 2 & 3 (see *Discussion*).

## Contract exercise

Now that we have a quantitative model of the mismatches between overruns caused and overruns received by individual stakeholders in a specific scenario, we move to the final piece of the analysis, in which we examine the consequences of these mismatching overruns under different contracting frameworks. The three contract types we examine are performance-based contracts (in which profits are tied to some performance milestones); fixed-price contracts (in which the stakeholder is awarded a fixed sum and must internalize any unexpected cost overruns); and cost-plus contracts (in which profits are awarded as a fixed percentage on top of realized costs).

Table 2 lays out the terms we chose to specify for each contract type. These terms are also visualized in the left column of Figure 5 (subfigures A, C and E). The contract terms establish the level of profits a given stakeholder is allowed to receive as a function of the overruns in their cost scope. As discussed in *Introduction*, “overruns in a stakeholder’s cost scope” are defined as any costs incurred by a stakeholder that go beyond the payments they would receive in the baseline case of the well-executed PWR12-BE construction cost model. The baseline cost scopes for the Construction Subcontractors, Design & Management, and Equipment Suppliers are \$614 million, \$325 million, and \$518 million in 1987 USD, respectively (see *Supplemental Information*).

**Table 2. Terms of three hypothetical contract structures used for the example in Figure 5.**

| | **Performance-Based** | **Fixed-Price** | **Cost-Plus** |
|---|---|---|---|
| **Profit Margin @ 30% Cost Overrun** | 8% | 8% | 8% |
| **Profit Margin @ 60% Cost Overrun** | 0% | −12.25% | 8% |
| **Maximum Profit Margin** | 16% | 40.40% | 8% |
| **Minimum Profit Margin** | 0% | −∞ | 8% |

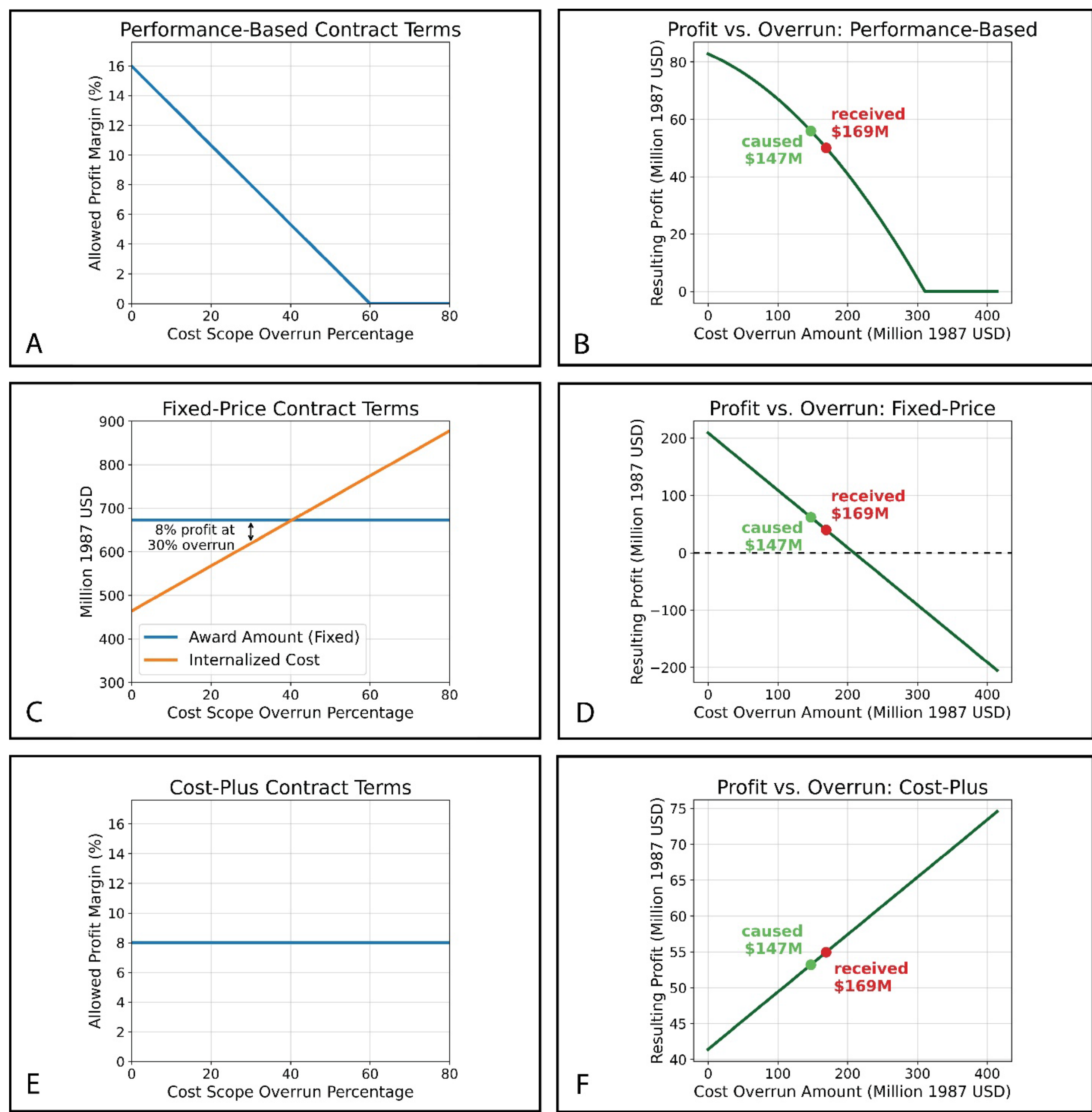

**Figure 5. Resulting profit outcomes for the Equipment Suppliers stakeholder in the modeled PWR12-BE / PWR12-ME scenario under three different contract structures.** Left column (A, C, E): terms of the three proposed contract structures. Right column (B, D, F): resulting profit vs. overrun curves for the Equipment Suppliers stakeholder under each of the three contract structures. Figures (A) and (B) depict a performance-based contract; (C) and (D) are for a fixed-price contract; and (E) and (F) are for a cost-plus contract. Light green points indicate the profit outcome for the Equipment Suppliers if profits are allocated based on the overruns they cause (i.e., the "cause-based allocation"), while red points indicate the outcome if profits are tied to the overruns the Equipment Suppliers receive payment for (i.e., the "recipient-based allocation").

For consistency, all three contract types are assumed to allow an 8% profit margin at 30% cost overruns – that is, a 30% 'contingency' is assumed, with an 8% profit margin allowed at this expected level of overrun. In the cost-plus case, the profit margin remains fixed at 8% regardless of the level of overrun in a stakeholder's cost scope. In the fixed-price case, the stakeholder's profit margin depends on how much the stakeholder's realized costs overrun their fixed award amount. In the performance-based case, the allowed profit margin varies linearly as a function of the overruns in a stakeholder's cost scope, with 16% profits allowed if there are no overruns, 8% profits allowed at 30% overruns, and 0% profits allowed at 60% overruns and beyond.

In the right column of Figure 5 (subfigures B, D and F), we apply these contract terms to the Equipment Suppliers stakeholder and illustrate the profits they can expect to receive as a function of the overruns in their $518 million baseline cost scope under each of the three contract types. (See *Supplemental Information* for details on how these profit vs. overrun curves are generated.) Our stakeholder analysis exercise established that the Equipment Suppliers caused 11.2% of the total $1.31 billion in overruns between the PWR12-BE and PWR12-ME ($147 million) and received 12.9% of the overruns as payment ($169 million), so we mark these two points on the profit vs. overrun curves. These points represent the profits the Equipment Suppliers stakeholder will receive if their responsibility for cost overruns is assessed based on the cost of overruns they charge to the project owner (red point, labeled "received $169M") or based on the amount of overruns they actually caused (green point, labeled "caused $147M").

The figures show that if the project owner chooses to assess the overrun amount attributable to the Equipment Suppliers stakeholder based on the excess costs the stakeholder receives as payment (which we will hereafter refer to as a "recipient-based allocation"), the stakeholder's realized profits can be significantly different compared to a scenario in which their profit margin is tied to the amount of overruns they are actually responsible for causing (hereafter referred to as a "cause-based allocation"). Under the recipient-based allocation, the Equipment Suppliers will receive $50.1M, $40.1M, and $54.9M in the performance-based, fixed-price, and cost-plus contract cases, whereas under the cause-based allocation, they will receive $56.0M, $62.0M, and $53.2M respectively. The differences between these allocation approaches is reported in Table 3 for the Equipment Suppliers, as well as for the Construction Subcontractors and Design & Management stakeholders.

**Table 3. Differences in profits received (in million 1987 USD) if profit allocations are recipient-based rather than cause-based for the Construction Subcontractors, Design & Management, and Equipment Suppliers under each of the three contract frameworks considered in Table 2.**

| | Performance-Based | Fixed-Price | Cost-Plus |
|---|---|---|---|
| **Construction Subcontractors** | (11) | (61) | 5 |
| **Design & Management** | 0 | 30 | (2) |
| **Equipment Suppliers** | (6) | (22) | 2 |

As Table 3 shows, stakeholder profits vary significantly depending on whether a cause-based allocation or a recipient-based allocation of profits is applied. This has important implications for contract structuring and administration, especially in the performance-based and fixed-price cases. For the performance-based case, it may seem natural to the owner of a nuclear construction project to assess the stakeholders' responsibility for cost overruns simply by looking at the amount of overruns they charge to the owner (i.e., a recipient-based allocation of profits). If this approach is used, the Equipment Suppliers will receive approximately $50.1 million (1987 USD) in profits based on the $169 million in overruns charged to the owner. In this scenario, however, the Equipment Suppliers only caused $147 million in overruns. If profits are tied to this amount instead (i.e., a cause-based allocation of profits), the Equipment Suppliers will receive approximately $56.0 million in profits. Thus, if the allowable profit margin is assessed using a recipient-based allocation, the Equipment Suppliers will experience a 10.5% decrease in profits compared to a cause-based allocation. In the fixed-price case, each stakeholder receives a fixed award amount, and any overrun in their cost scope will eat into their profit margin, even if these overruns were caused through no fault of their own. Table 3 shows that the Construction Subcontractors and the Equipment Suppliers lose about $61 million and $22 million respectively due to overruns caused by Design & Management and the Regulator under the fixed-price contract.

These profit discrepancies are likely to be noticed, and the Equipment Suppliers and Construction Subcontractors may decide to sue the stakeholders responsible for causing the additional cost overruns – in this case, Design & Management and perhaps even the Regulator – to recover their lost profits. Such disputes between stakeholders can result in misaligned incentives, further schedule delays, and further cost overruns. This is not merely a theoretical concern: in the real world, expensive and time-consuming lawsuits over lost profits due to unfair attribution of overrun responsibility happen often in construction megaprojects [18]. If performance-based and fixed-price contracts are not designed to account for this, the companies involved will try to force accountability through litigation.

In the cost-plus case, the Equipment Suppliers and Construction Subcontractors will earn a smaller profit under a cause-based allocation compared to the recipient-based allocation, which is the opposite of the fixed-price and performance-based cases. This is because these two stakeholders both received payment for more overruns than they caused (see Figure 3), and the slope of the profit-vs-overrun curve is positive for the cost-plus contract, whereas the slope is negative for each of the other two contract types (see Figure 5). For stakeholders that caused more overruns than they received payment for, the opposite will be true, as shown in Table 3 for the Design & Management stakeholder.

Table 3 also shows that the fixed-price contract tends to result in the most extreme profit deltas, which suggests a greater likelihood of lawsuits between stakeholders to recover unfairly lost profits. It is also noteworthy that the profit delta is zero for Design & Management in the performance-based case. This is because this stakeholder has very large cost overruns (causing $532M and receiving $501M), which means they end up in the zero-profit region of the profit vs. overrun curve (see Figure 5(B)) regardless of whether their profits are tied to overruns received or overruns caused.

The cost-plus case is unique because both the Construction Subcontractors and the Equipment Suppliers are better off if their profits are based on a recipient-based allocation rather than a cause-based allocation. The Design & Management stakeholder loses $2 million if this approach is used (see Table 3), but under a cost-plus arrangement, there is no inherent reason why profits should be tied to overruns caused rather than to overruns received. Therefore, in the cost-plus case, all three stakeholders are less likely to sue each other than in the fixed-price or performance-based cases, as all parties will always receive an 8% profit margin on the costs they charge to the owner. The downside, of course, is that this contract structure creates no incentive for the stakeholders to minimize the overruns they charge the project owner for, which could result in higher project costs and potentially litigation between the owner and these stakeholders. Clearly, all three contracting styles have advantages and disadvantages that must be carefully weighed at the project outset.

## Discussion

The results of the analysis show that profit outcomes for individual stakeholders can vary significantly depending on whether profits are awarded using a recipient-based allocation or a cause-based allocation, which can potentially result in costly and time-consuming litigation between parties, especially in the performance-based and fixed-price contract structures. When a recipient-based allocation is used in the performance-based or fixed-price arrangement, lawsuits may be filed by stakeholders who are being punished for overruns in their cost scope that were in fact caused by other stakeholders.

For example, consider the Vogtle 3 & 4 project, in which a fixed-price contracting structure was used to compensate contractors for excavation work. Westinghouse and Stone & Webster billed the owners, a consortium of utilities in the Southeastern U.S., $58 million in extra charges above the original price for excavation work, alleging that

the owners had provided incorrect and incomplete information about site conditions when the contractors prepared their original bid [5]. The owners paid half of this amount but disagreed with the contractors about their responsibility for the full $58 million in overruns, and the contractors sued to recover the remainder. As of 2012, before the first nuclear-grade concrete had been poured at Vogtle 3 & 4, the excavation lawsuit was reportedly only “the latest in a series of disputes” between the owners and the contractors, including an “$800 million dispute stemming from delays in getting key regulatory approvals” and additional disputes over design change orders, “some of which have gone unresolved for months” [5]. A 2015 report by Bechtel on the cost and schedule overruns at the V.C. Summer 2 & 3 project similarly found that “the Contract does not appear to be serving the Owners or the Consortium particularly well” [4], with expert testimony revealing disagreements between Westinghouse, Chicago Bridge & Iron, and the project owners about which party was responsible for excess costs [3].

The analysis suggests that cost-plus frameworks may be a superior choice for avoiding litigation between stakeholders directly involved in delivering the project (Construction Subcontractors, Equipment Suppliers, Design & Management) due to the fixed profit margin. On the other hand, cost-plus contracts can also increase project costs by incentivizing cost overruns in all stakeholders’ project scopes, thereby making their incentive structures misaligned with the project owner. Taken together, these results suggest that regardless of the contract type chosen, a high level of active owner engagement is critical for minimizing project costs. In the case of a cost-plus contract, the task of the owner is straightforward: build up a sizeable internal project team to carefully monitor the day-to-day activities of each stakeholder involved in project delivery and make sure all costs they charge to the owner are justified. For fixed-price and performance-based contracts, this powerful oversight role is also important, but there should also be a greater emphasis on ensuring that profit outcomes for each stakeholder are always cause-based rather than recipient-based. The goal of the owner, in other words, should be to ensure that each stakeholder is rewarded fairly for their performance, and never punished for overruns caused by other parties. Otherwise, the stakeholders will take it upon themselves to force accountability through litigation, potentially escalating the cost and schedule of the project and making stakeholders less incentivized to collaborate productively on the worksite.

As an example of how project owners might structure a contract to ensure that stakeholder performance is always assessed accurately, consider a performance-based contract in which profits are tied to meeting certain construction milestones within a specified time window. Instead of specifying a date for completion, an approach commonly used for milestone structuring (e.g., “if subcontractor X installs the containment building roof by August 12, they will receive a 5% bonus in compensation”), the contract could establish milestones with a set of firm preconditions for the completion of other stakeholders’ work scopes. For example, “upon completion of the containment walls by subcontractor Y and all open-top equipment installation activities by subcontractor Z, if subcontractor X then installs the containment building roof within 7 days, they will receive a 5% bonus in compensation". This could help to avoid situations in which one stakeholder's delays impede another stakeholder's ability to meet their contractual milestones, which could result in subcontractor X suing subcontractor Y or Z over lost profits. Using this approach, project participants can avoid any potential confusion over who is responsible for causing overruns, and compensating stakeholders fairly based on their performance may become more straightforward. An alternative, hybridized approach between "date-based" and "time-window-based" schedule milestones that has been used in real-world construction projects and may be easier to administer is to continuously establish a series of near-term milestone dates on a "rolling basis" over the lifetime of the construction project [19]. Near-term milestone dates may narrow the window of opportunity for one stakeholder to unexpectedly impede the progress of another, while also requiring less information about the completion preconditions of other work activities. Furthermore, the inability to accurately forecast all potential schedule slippage on the critical path of a nuclear construction project means that milestone dates established exclusively at the project outset may "lose credibility with the contractor[s]" [19].

While attention to detail in the initial contract drafting is crucial, the reality is that even the most careful planners will not be able to anticipate every dispute or unfair allocation of profits that may arise during construction. Thus, the owner should also set up an internal team of project monitors to carefully oversee the day-to-day activities of all stakeholders involved in construction and should be prepared to step in and offer acceptably revised contract terms if a dispute between parties appears imminent. This effort may be assisted by establishing an internal conflict resolution board, which creates a setting for stakeholders to try and resolve disputes productively without resorting to litigation. Flyvbjerg and Gardner provide an illustrative example of the successful application of flexible contract terms during the construction of Heathrow Airport's Terminal Five (T5) [2]. During the Heathrow project, a steel-frames contractor operating under a fixed-price contract faced delays due to a concrete-pouring contractor falling behind schedule, thereby threatening the steel contractor's profit margin. To mitigate the risk of litigation or project abandonment by the steel contractor, the project owner, British Airports Authority (BAA), intervened and transitioned the steel contractor to a cost-reimbursable contract arrangement with an additional percentage profit contingent upon meeting certain milestones. This resulted in a more cooperative, problem-solving dynamic between the two contractors, wherein the concrete contractor agreed to increase the workforce and the steel contractor adjusted its workflow to accommodate the delays. The resolution of this potentially disruptive conflict between two stakeholders allowed the project to progress more smoothly, reducing the risk of further cost and schedule overruns.

It is important to note that flexible contracting and a greater level of owner participation come with increased owner's costs, which represented ~14% of total project OCC at Vogtle 3 & 4 [20]. This figure would be further increased by costs associated with more extensive pre-project planning, establishing an internal dispute resolution board, hiring a large internal staff to actively oversee day-to-day construction activities, and tweaking contract terms to increase the compensation of one or more stakeholders. The question is therefore whether the costs associated with this approach outweigh the benefits. While our analysis does not model the additional owner's costs associated with increased stakeholder engagement or attempt to quantify the specific cost and schedule savings from averting conflicts between stakeholders, findings from previous authors support the conclusion that greater owner involvement does typically reduce megaproject costs. As discussed in *Introduction*, some previous econometric studies of nuclear power plant capital costs have found that costs are lower for projects in which the owner acts as its own construction manager [10] [11]. There also seems to be general agreement in the qualitative literature on megaproject management that better pre-project planning and more proactive owner engagement with stakeholders tends to result in better project outcomes [2] [12] [13], and the 2024 DOE report *Pathways to Commercial Liftoff: Advanced Nuclear* has highlighted the importance of more extensive project management and increased spending on pre-project planning for FOAK projects [21].

Thus, the results of our analysis reinforce the conclusions of other authors who have found that detailed pre-project planning and more active engagement by the project owner can help to reduce cost and schedule overruns in nuclear power plant construction. Using our novel approach to stakeholder analysis, we are able to model the stark differences in financial outcomes that can result when contracts tie stakeholder profits to cost overruns received rather than cost overruns caused. For modern nuclear construction projects costing tens of billions of dollars, using a recipient-based rather than cause-based profit allocation mechanism to administer the contract could potentially result in profit misallocations on the order of hundreds of millions to billions of dollars, which helps to explain why incentive misalignments and lawsuits between project participants so often arise in multi-stakeholder nuclear power plant construction projects. These results also suggest that one reason the vertically integrated project delivery structures used in Russia, China, Japan, and South Korea tend to achieve better outcomes than Western nuclear builds [6] is that the existence of fewer unique stakeholders creates fewer opportunities for incentive misalignment, profit misallocations, and resulting overruns. Our conclusions enable a better understanding of the importance of careful contracting and proactive owner involvement in Western nuclear projects with many distinct stakeholders involved.

## Limitations

We emphasize that our analysis is primarily a stylized exercise intended to show how overrun mismatches can affect project stakeholders under a specific set of circumstances. We rely on some empirical data from past nuclear projects to inform the structure of our stakeholder analysis model (e.g., a 1980 survey of nuclear construction productivity [15]) and our choice of scenario inputs (e.g., productivity data from the Vogtle 3 & 4 experience [17]), but the scenario we model is not based on a single set of empirical data from any real-world nuclear construction project. In fact, the PWR12-BE and PWR12-ME reactor cost models used as the foundation of the analysis were created by United Engineers and Constructors [8] to be representative of typical construction experience in the late 1980s, and do not reflect the economics of any individual reactor project. The equations used to model overrun mechanisms are calibrated to sum to the total OCC of the PWR12-ME, but the breakdown of costs into site materials, site labor, factory equipment, and indirect costs will vary depending on the scenario modeled. For the inputs chosen in Table 1, the synthetic OCC amounts are $146 million in site materials, $490 million in site labor, $687 million in factory equipment, and $1.207 billion in indirect costs, which differs from the baseline PWR12-ME model in EEDB Phase IX [8] by -3.6%, -2.9%, +29.3%, and -10.1% respectively.

We also do not consider the exhaustive list of all possible overrun mechanisms in our analysis, choosing to focus only on rework, low labor productivity, and supply chain delays. Other overrun mechanisms might include exogeneous factors such as market impacts on labor and commodity prices, or political or legal delays driving longer schedules. Our choice of which stakeholders to model (see Figure 1) is somewhat contrived based on the structure of the EEDB code of accounts and the structure of INL's Cost Reduction Tool equations for overruns (see Methods), and there is no guarantee that work scopes would be broken down neatly between the five stakeholders we consider in a real-world project, but the takeaways are still generally applicable to a Western-style project with multiple stakeholders involved in project delivery. This is not a complete list of all limitations to our approach, but should provide a more precise picture of the novelty and value of our analysis.

## Future work

The results presented in this work represent a single modeling scenario using the selected input values presented in *Results § Input assumptions*, but modeling additional scenarios with different stakeholder proficiency levels, labor productivity values, and financing rates could potentially be valuable for understanding the magnitude of overrun mismatches and profit misallocations that can arise in a wide variety of circumstances. Applying real-world contract frameworks with more detailed and complex provisions than those used in Table 2 (e.g., time-based milestones for specific deliverables) could also be useful to pursue in future work. Subdividing the five stakeholders examined in this analysis into more subgroups (e.g., modeling specific subcontractors within the Construction Subcontractors stakeholder, capturing the individual effects of the Reactor Vendor, A-E, and Constructor within the Design & Management team, and modeling the different repayment structures for banks, equity investors, etc. within the Creditors stakeholder) would provide a greater level of granularity to the analysis. Lastly, given that a primary recommendation of this study is a greater level of owner engagement and oversight in nuclear construction, it would be valuable to attempt to directly quantify owner's costs (and owner's benefits) associated with a greater level of active owner involvement in the project. A cost-benefit analysis of owner engagement could be performed to determine an ideal level of in-house staffing to ensure project success.

## Key takeaways

We conclude with a summary of four key recommendations about the approaches new nuclear power plant stakeholders should take to minimize costs and maximize the chances of project success.

(1) **Increased attention to detail in contract structuring at the project outset.** For contracts in which rewards are tied directly to performance (e.g., fixed-price or performance-based contracts), careful structuring is essential to ensure that all stakeholders are rewarded fairly for their performance and never punished unintentionally for overruns or delays caused by other parties. The *Discussion* section provides a few specific examples of how contracts might be structured to ensure these outcomes, including establishing well-defined preconditions for subcontractor work scopes to begin and using "time-window-based" instead of "date-based" schedule milestones.

(2) **Ongoing owner engagement with project stakeholders.** As the project moves from the planning stage to the execution stage, ongoing owner involvement is crucial to keep the project on track. For cost-plus contracts, the owner should build up a sizeable and well-informed oversight team to make sure all expenses are justified. For fixed-price and performance-based contracts, the owner should also take additional steps to ensure that all stakeholders are compensated fairly for their performance as conditions on the ground change. The *Discussion* section suggests flexible contract terms and internal dispute resolution boards as two possible steps towards achieving this goal. If the project owner does not take action to ensure performance is rewarded accurately, the result will be a dealignment of incentives and lawsuits between stakeholders to recover lost profits, as has occurred often in recent Western builds.

(3) **More time and resources spent on planning and oversight.** Previous studies have highlighted extensive pre-project planning and strong owner oversight as two of the strongest factors in reducing megaproject costs and schedule risks. As the recent 2024 DOE Liftoff report on advanced nuclear [21] and an IAEA report on nuclear construction management [22] have highlighted, as much as half of a nuclear construction project timeline should consist of pre-project planning. Although owner's costs already represent ~10-20% of OCC for nuclear projects [20] [23] [24], additional spending of a few percentage points in this category would be well justified if it can help to mitigate avoidable cost overruns due to rework, low productivity, supply chain delays, and associated financing costs, especially for first-of-a-kind projects where these overruns may represent 50% or more of total project spending.

**(4) Vertical integration.** The takeaways above are primarily intended for nuclear construction projects involving many independent stakeholders (owner, construction manager, architect-engineer, equipment suppliers, reactor vendor, etc.) whose incentives must be forcibly aligned using carefully structured contracts. Another approach that has been shown to deliver nuclear construction projects effectively is a vertically integrated model in which most or all of these functions are performed by a single entity with internally aligned incentives, often a national government through the use of state-owned enterprises, as in the French, Russian, Chinese, Indian, and South Korean nuclear industries. In countries where a state-owned model is not feasible, vertical integration within private companies is also a potential option for minimizing the number of stakeholders and reducing opportunities for incentive misalignment. Historical examples have included large utility companies in the U.S. and Canada (Ontario Hydro, Commonwealth Edison, Duke Power, Tennessee Valley Authority, etc.) taking on most or all of the engineering, procurement, and construction (EPC) responsibilities for their plants internally [10], as well as the historical German and Japanese models in which the owner and EPC are separate entities but most EPC functions are handled by a single vertically integrated company (Hitachi, Toshiba, Kraftwerk Union, etc.) [25]. Lastly, given the critical importance of incentive alignment, if government incentive programs with limited resources (loan guarantees, tax credits, direct investments, cost overrun insurance, etc.) are faced with multiple applications for new nuclear projects, all else being equal, administrators should consider prioritizing support for merchant projects in deregulated electricity markets over cost-of-service projects in regulated markets. Prioritizing projects in which the incentives of the project owner are aligned as much as possible with end-use consumers may increase the chance of success for these government programs.

# Methods

As discussed in *Introduction*, our goal is to create a stylized scenario to analyze stakeholder involvement in nuclear construction cost and schedule overruns. In order to create a scenario highlighting the effects of these cost overrun mismatches under different contracting frameworks, we begin by considering two bottom-up cost models for nuclear power plant construction from United Engineers & Constructors' Energy Economic Data Base (EEDB) Phase IX report: the PWR12-ME (Median Experience), which reflects the median construction experience for an 1144-MWe pressurized water reactor entering service in the U.S. in 1987, and the PWR12-BE (Better Experience), which reflects costs for "a small group of single unit nuclear power plants at the low end of the range of base construction costs for single unit nuclear power plants currently entering service" in 1987 [8]. We take the PWR12-BE to be our cost model for a well-executed nuclear construction project, and the PWR12-ME to be our cost model for a poorly-executed nuclear construction project after accounting for cost and schedule overruns.

Since the EEDB models only include overnight capital costs (OCC), we also add in financing costs to better capture the full cost of a nuclear construction project. To calculate the cost of interest during construction (IDC), we assume OCC accumulates over the duration of the project according to a sinusoidal spending curve, and IDC accumulates on the OCC until the project is complete, according to the equations in *Supplemental Information*. As discussed in *Results* § *Input assumptions*, we assume an effective interest rate of 4% for our analysis.

The next few sections describe how we approached each of the four steps of the analysis listed in *Introduction* § *Analytical approach* to model:

(1) How the overruns between the PWR12-BE and ME arose mechanistically (*Overrun mechanisms*);
(2) Which stakeholders caused the overruns (*Overruns by causer*);
(3) Which stakeholders received payment for the overruns (*Overruns by recipient*);
(4) How mismatches between overruns caused and overruns received by each stakeholder affect the stakeholders' profit margins under different contract frameworks (*Contract analysis*).

## Overrun mechanisms

For this step, we borrow relations from INL's Cost Reduction Tool [14], which introduced simple correlations for modeling cost overrun mechanisms to determine how nuclear construction costs might reduce across multiple reactor deployments under different circumstances. For our analysis, we use equations from the CRT to model three specific mechanisms for OCC and schedule overruns between the PWR12-BE and PWR12-ME – rework, low construction site productivity, and supply chain delays – and we also capture financing cost overruns caused by these sources of OCC and schedule overruns.

**Rework**

Overruns due to rework may be caused by human errors on the worksite that require work to be redone, or when a change order is issued by the design and management team [1]. As described in Appendix C of the original 2024 report on the Cost Reduction Tool [14], the CRT uses a "total rework factor," $R_{tot}$, to calculate the impact of rework on costs:

$$\Delta C_{direct} = (R_{tot} - 1) \times \left(C_{direct,material} + C_{direct,factory}\right) + \left[\frac{R_{tot}}{p} - 1\right] \times C_{direct,labor} \quad \text{(Equation 1)}$$

Here, $\Delta C_{direct}$ refers to the change in direct costs between the PWR12-BE and PWR12-ME. "Direct costs" include structures and improvements (Account 21 in the EEDB code of accounts), reactor plant equipment (Account 22), turbine plant equipment (Account 23), electric plant equipment (Account 24), miscellaneous plant equipment (Account 25), and the main condenser heat rejection system (Account 26). In EEDB's PWR12-BE and ME cost models, costs are subdivided into site material costs, site labor costs, and factory equipment costs, which are indicated by the subscripts *material*, *labor*, and *factory* in Equation 1.

The total rework factor is formulated in the Cost Reduction Tool as the product of three separate rework factors:

$$R_{tot} = R_{AE} \times R_C \times R_{design} \quad \text{(Equation 2)}$$

Where $R_{AE}$, $R_C$, and $R_{design}$ reflect rework caused by a lack of architect-engineer (A-E) proficiency, a lack of construction subcontractor proficiency, and the reactor design not being fully complete prior to construction start, respectively. In the Cost Reduction Tool, the user specifies a proficiency value between 0 and 2 for the A-E and the construction subcontractors, and $R_{AE}$ and $R_C$ are calculated as a function of this proficiency value. (These proficiency values are referred to as "levers" in the Cost Reduction Tool [14].) For the scenario we model with the PWR12-BE and ME, the equations are:

$$R_{AE} = -0.1266 \times AEP + 1.253 \quad \text{(Equation 3)}$$

$$R_C = -0.1266 \times CSP + 1.253 \quad \text{(Equation 4)}$$

Where $AP$ and $CSP$ are the A-E proficiency lever and the construction service proficiency lever, respectively. Since the levers take values between 0 and 2, $R_{AE}$ and $R_C$ may range from 1 to 1.253 depending on the proficiency of the stakeholders. For a description of how the coefficients in Equations 3 and 4 are derived for the scenario we model in this analysis, see *Supplemental Information*.

Equation 1 shows how direct cost overruns due to rework are calculated. Throughout our analysis, whenever overruns in direct costs are calculated, overruns in indirect costs

– including construction services (Account 91), engineering & home office services (Account 92), and field supervision & field office services (Account 93) – must also be calculated. See *Supplemental Information* for a description of the equations we used to scale indirect costs with direct costs.

In addition to costs, the change in construction duration as a result of rework must be calculated. We adopt the following equation from the Cost Reduction Tool:

$$\frac{\text{new construction duration}}{\text{baseline construction duration}} = 0.3 \times \frac{R_{tot}}{p} + 0.7 \qquad \text{(Equation 5)}$$

As noted in Appendix C of the original CRT report [14], this equation comes from a cost modeling exercise of multiple reactor architectures in which Stewart and Shirvan observed that when total labor hours increased by a factor of ~5.4, total construction duration increased by a factor of ~2.2 [26].

**Low Productivity**

Part of the reason for the higher cost of the PWR12-ME compared to the PWR12-BE is lower labor productivity at the construction site. For our illustrative scenario, we assume the labor productivity of the PWR12-ME is 75% of the PWR12-BE productivity. This choice is informed by difficult construction experience at Vogtle 3 & 4, in which the "Cost Performance Index" ($CPI$ = hours spent / hours earned) was reported to be ~1.3 - 1.4 as of 2020-21 [16] [17], corresponding to a labor productivity of $1 / \mathrm{CPI} \approx 0.75$. This productivity value is plugged into Equation 1 (alongside $R_{tot}$ for rework) to determine how direct costs increase with low productivity, and changes in indirect costs and construction duration due to low productivity are calculated using Equations S7 - S10 and Equation 5, respectively.

**Supply Chain Delays**

The total construction duration overrun between the PWR12-BE and PWR12-ME is 98 months - 72 months = 26 months [8]. Equation 5 gives the change in construction duration due to rework and low productivity, which, with the inputs we select for our analysis (see Table 1), account for ~16.6 of the total 26 months of schedule overrun. We assume that the remaining 9.4 months of schedule overruns are caused by supply chain delays. As noted in *Limitations*, rework, low productivity, and supply chain delays are not the only possible sources of schedule overruns in real-world nuclear construction projects, but they are the only three mechanisms we consider for the sake of our analysis.

The change in construction duration due to supply chain delays also affects indirect OCC due to the appearance of construction duration in Equations S9 - S10. However, the impact of supply chain delays on OCC is relatively minimal, accounting for only 4.9% of the total OCC overruns in the scenario we model.

## Overruns by causer

Having established four mechanisms by which overruns can occur (rework, low productivity, supply chain delays, and financing escalation directly related to these sources of OCC and schedule overruns), we next needed to establish a plausible framework to model which of the five stakeholders were responsible for causing each of these four overrun types.

**Rework by Causer**

To assign how much of the rework-related OCC overruns were caused by each stakeholder, the following expressions were used:

$$\Delta C_{rework,C} = \Delta C_{rework,tot} \times \frac{\Delta C_{rework}(R_C, 1,1)}{\Delta C_{rework}(R_C, 1,1) + \Delta C_{rework}(1, R_{AE}, 1) + \Delta C_{rework}\left(1,1, R_{design}\right)} \quad \text{(Equation 6)}$$

$$\Delta C_{rework,AE} = \Delta C_{rework,tot} \times \frac{\Delta C_{rework}(1, R_{AE}, 1)}{\Delta C_{rework}(R_C, 1,1) + \Delta C_{rework}(1, R_{AE}, 1) + \Delta C_{rework}\left(1,1, R_{design}\right)} \quad \text{(Equation 7)}$$

$$\Delta C_{rework,design} = \Delta C_{rework,tot} \times \frac{\Delta C_{rework}\left(1,1, R_{design}\right)}{\Delta C_{rework}(R_C, 1,1) + \Delta C_{rework}(1, R_{AE}, 1) + \Delta C_{rework}\left(1,1, R_{design}\right)} \quad \text{(Equation 8)}$$

Where $\Delta C_{rework,tot}$ is the total rework OCC overrun, and $\Delta C_{rework,C}$, $\Delta C_{rework,AE}$, and $\Delta C_{rework,design}$ are the rework cost overruns caused by lack of construction proficiency, lack of A-E proficiency, and lack of design completion, respectively. $\Delta C_{rework}(R_C, 1,1)$ represents the value of $\Delta C_{rework,tot}$ calculated using Equations 1 and 2 with $R_C$ taking on its default value for our analysis (1.063, see Table 1), $R_{AE}$ and $R_{design}$ each being set to 1 implying no rework due to lack of A-E proficiency or design incompletion (see *Overrun mechanisms* § *Rework)*, and other inputs to the analysis in Table 1 being held constant ($p$ = 0.75, $r$ = 4%). The values of $\Delta C_{rework}(1, R_{AE}, 1)$ and $\Delta C_{rework}\left(1,1, R_{design}\right)$ are defined similarly. Responsibility for rework-related schedule overruns is calculated using the same relations as for rework-related cost overruns, but with schedule overruns ($\Delta T$) substituted for cost overruns ($\Delta C$) in Equations 6 - 8.

We attribute $\Delta C_{rework,C}$ to the Construction Subcontractors, and $\Delta C_{rework,AE}$ to Design & Management, a high-level stakeholder that includes within it the A-E (see Figure 1). Since the 1980s EEDB cost models are from an era when design changes were frequently mandated by the U.S. Nuclear Regulatory Commission [9] [27], we attribute all responsibility for $\Delta C_{rework,design}$ to the Regulator for the sake of our illustrative analysis. This results in the Regulator being responsible for 36.1% of the total rework OCC overruns and 22.2% of the overall cost overruns in the scenario we construct here, but other breakdowns are possible depending on the choice of inputs for the architect-engineer proficiency ($AEP$) and construction service proficiency ($CSP$) levers (see Table 1). In reality, the Design & Management stakeholder would likely be responsible for some portion of the rework due to an incomplete design prior to construction start, but we try to capture their influence on rework overruns through the $AEP$ lever.

**Low Productivity by Causer**

To model which stakeholders caused the most productivity losses, we used data from a historical survey of U.S. nuclear power plant construction workers across multiple sites [15]. The survey reported the relative shares of unproductive weekly working hours attributed by workers to six different sources: material availability, tool availability, crew interfacing, overcrowded work areas, instructions time, and inspection delays.

The results of the construction worker survey do not immediately reveal how responsibility for unproductive working hours is distributed among the handful of stakeholders defined in our analysis. To address this, we started by determining which stakeholders could have caused each of the six categories of low productivity. This information can be used to determine the minimum and maximum possible responsibility of each stakeholder for non-rework-related unproductive hours. A description of how the six categories of low productivity were mapped to each stakeholder is included in *Supplemental Information*.

Since we do not know exactly where each stakeholder falls within their potential range of responsibility for low productivity, we simply assume the midpoint value for each stakeholder, as illustrated in Figure 6.

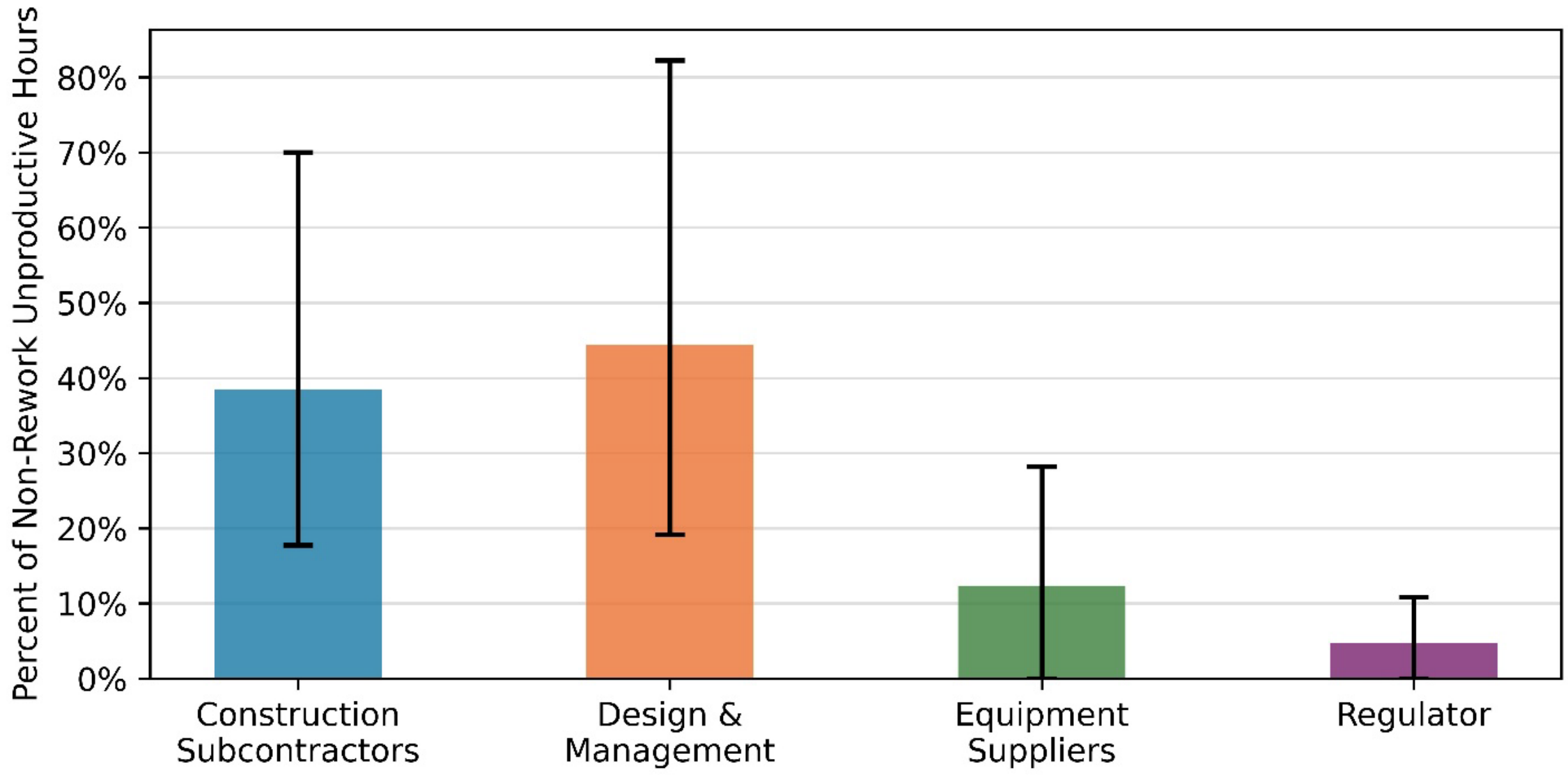

**Figure 6. Assessed responsibility of each stakeholder for low productivity-related overruns.** Black error bars show the range of possible responsibility for low labor productivity by stakeholder. Colored bars represent midpoint values of responsibility. Note that these four midpoint values initially added up to 114% and had to be re-normalized to sum to 100%, as described in Supplemental Information, so the colored bars do not fall at the exact midpoint of the black error bars.

Once each stakeholder's midpoint responsibility for non-rework unproductive hours was determined, their corresponding share of responsibility for productivity-related overruns was calculated using Equation 9:

$$\Delta C_{LP,i} = \Delta C_{LP,tot} \times f_{LP,i} \quad \text{(Equation 9)}$$

Where $\Delta C_{LP,i}$ is the low-productivity cost overrun caused by stakeholder $i$, $\Delta C_{LP,tot}$ is the total low-productivity cost overrun (obtained using the same methodology as described in the original report on the Cost Reduction Tool [14]), and $f_{LP,i}$ is the midpoint percentage of non-rework unproductive hours (normalized to 100%) caused by stakeholder $i$, shown in Figure 6. Responsibility for low productivity-related schedule overruns is calculated using the same relations as for low productivity-related cost overruns, but with schedule overruns ($\Delta T$) substituted for cost overruns ($\Delta C$) in Equation 9.

**Supply Chain Delays by Causer**

OCC and schedule overruns related to supply chain delays were simply attributed to the Equipment Suppliers stakeholder. In reality, additional stakeholders may be the root causers of certain equipment delivery delays – for example, the Design & Management team may issue an unexpected change order for a component during fabrication. For future work, if empirical data on supply chain delays by causer can be obtained from historical nuclear builds, this information should be integrated into the analysis.

**Financing Overruns by Causer**

Since financing costs are a function of both the OCC and the project duration (see *Supplemental Methods*), the calculation of financing cost overruns caused by each stakeholder needs to take into account both the OCC overruns and the schedule delays caused by the stakeholder. This was achieved using Equation 10:

$$(\%\ \Delta C_{fin})_i = \frac{\Delta C_{fin,i}}{\sum_{i=1}^{n} \Delta C_{fin,i}} \times 100\% \quad \text{(Equation 10)}$$

In Equation 10, $\Delta C_{fin}$ is the total financing overrun for the project (calculated as explained in *Supplemental Methods*), $(\%\ \Delta C_{fin,tot})_i$ is the percentage share of $\Delta C_{fin}$ caused by stakeholder $i$, and $\Delta C_{fin,i}$ is defined as:

$$\Delta C_{fin,i} \equiv C_{fin}(OCC_i, T_i) - C_{fin}(OCC_0, T_0) \quad \text{(Equation 11)}$$

Here, $C_{fin}(OCC_0, T_0)$ is the baseline financing cost for the PWR12-BE ($175 million in 1987 USD), calculated by plugging the PWR12-BE overnight capital cost ($OCC_0$ = $1.456 billion in 1987 USD) and construction duration ($T_0$ = 72 months) into Equation S5 using the presumed interest rate of $r$ = 4% from Table 1. $C_{fin}(OCC_i, T_i)$ is also calculated using Equation S5, but $OCC_i$ and $T_i$ are calculated using Equations 12 and 13:

$$OCC_i = OCC_0 + \Delta OCC_i \quad \text{(Equation 12)}$$

$$T_i = T_0 + \Delta T_i \quad \text{(Equation 13)}$$

Where $\Delta OCC_i$ and $\Delta T_i$ are the OCC and schedule overrun caused by stakeholder $i$, calculated as explained in the preceding sections.

## Overruns by recipient

Because the EEDB reactor cost models subdivide costs into factory equipment, site material, and site labor costs [8], modeling which fraction of the cost overruns are received by each stakeholder as payment is relatively straightforward. For direct costs (Account 2), all factory equipment costs are assumed to go to the Equipment Suppliers stakeholder, while all site material and site labor costs go to the Construction Subcontractors stakeholder.

In EEDB, indirect costs (Account 9) are split into three subaccounts: Construction Services (Account 91), Engineering and Home Office Services (Account 92), and Field Supervision and Field Office Services (Account 93). Of the total $791M (1987 USD) in indirect cost overruns between the PWR12-BE and PWR12-ME, $185M (23.4%) of the overrun was in Account 91, while the remaining 76.6% was in Accounts 92 and 93. Since Account 91 includes temporary construction facilities, construction tools and equipment, and additional costs to constructors including payroll insurance, taxes, permits, and insurance, we assume 23.4% of all indirect cost overruns are received by the Construction Subcontractors (i.e., all costs from Account 91). Accounts 92 and 93 include engineering services, management services, supervision, testing and startup services, and field QA/QC, which are performed by some combination of the reactor vendor, constructor, architect-engineer, and possibly additional stakeholders (e.g., engineering or management consultants brought in on the project), but unfortunately the EEDB data does not subdivide these accounts in a way that would allow a determination of the percentages that go to each of these individual stakeholders. Instead, we attribute the entire 76.6% to a stakeholder called Design & Management (i.e., all costs from Accounts 92 and 93), which includes within it the reactor vendor, A-E, and constructor, as shown in Figure 1. Lastly, all financing overruns are assumed to be received by the Creditors, and we assume the Regulator does not receive any payment for overruns.

## Contract analysis

For the final piece of our analysis, we take the results of the cost and schedule overrun attributions from the preceding sections and use them to analyze how the profit margins of each stakeholder (Design & Management, Construction Subcontractors, Equipment Suppliers, Creditors, and Regulator) are affected by the mismatches between overruns caused and overruns received by each stakeholder. As discussed in *Results* (Figure 5), we perform this analysis under three different contract frameworks used commonly in construction megaprojects: performance-based contracting (in which profits are tied to some performance benchmark), fixed-price contracting (in which the stakeholder is awarded a fixed sum and must internalize any unexpected cost overruns), and cost-plus contracting (in which profits are awarded as a fixed percentage on top of realized costs).

The logic behind our choice of specific contracting terms is discussed in detail in *Results* § *Contract exercise*. The process used to generate "profit vs. overrun curves" (Figure 5, right column) from the contract terms (Figure 5, left column) is described in detail in *Supplemental Information*.

# Funding Acknowledgement

This material is based upon work funded by the U.S. Department of Energy (DOE) through the Systems Analysis & Integration (SA&I) Campaign and a University Nuclear Leadership Program (UNLP) Graduate Fellowship from the DOE Office of Nuclear Energy. Any opinions, findings, conclusions, or recommendations expressed in this publication are those of the authors and do not necessarily reflect the views of the DOE or the Office of Nuclear Energy.

# Supplemental Information

## Supplemental Methods

### Financing costs

As mentioned in *Methods*, the financing costs in this analysis are calculated using a sinusoidal spending curve. The cumulative fraction of OCC expended in year $t$ of construction ($F_t$) is given by:

$$F_t = \frac{1}{2}\left(1 - \cos\left(\frac{\pi t}{T}\right)\right) \quad \text{(Equation S1)}$$

Where $T$ is the construction duration in years. The resulting OCC spend curve is shown in Figure S1 for the PWR12-BE with a 72-month construction duration. This simple sinusoidal curve approximates the more detailed empirical sinusoidal curves used by authors including Mooz [S1] and Komanoff [9].

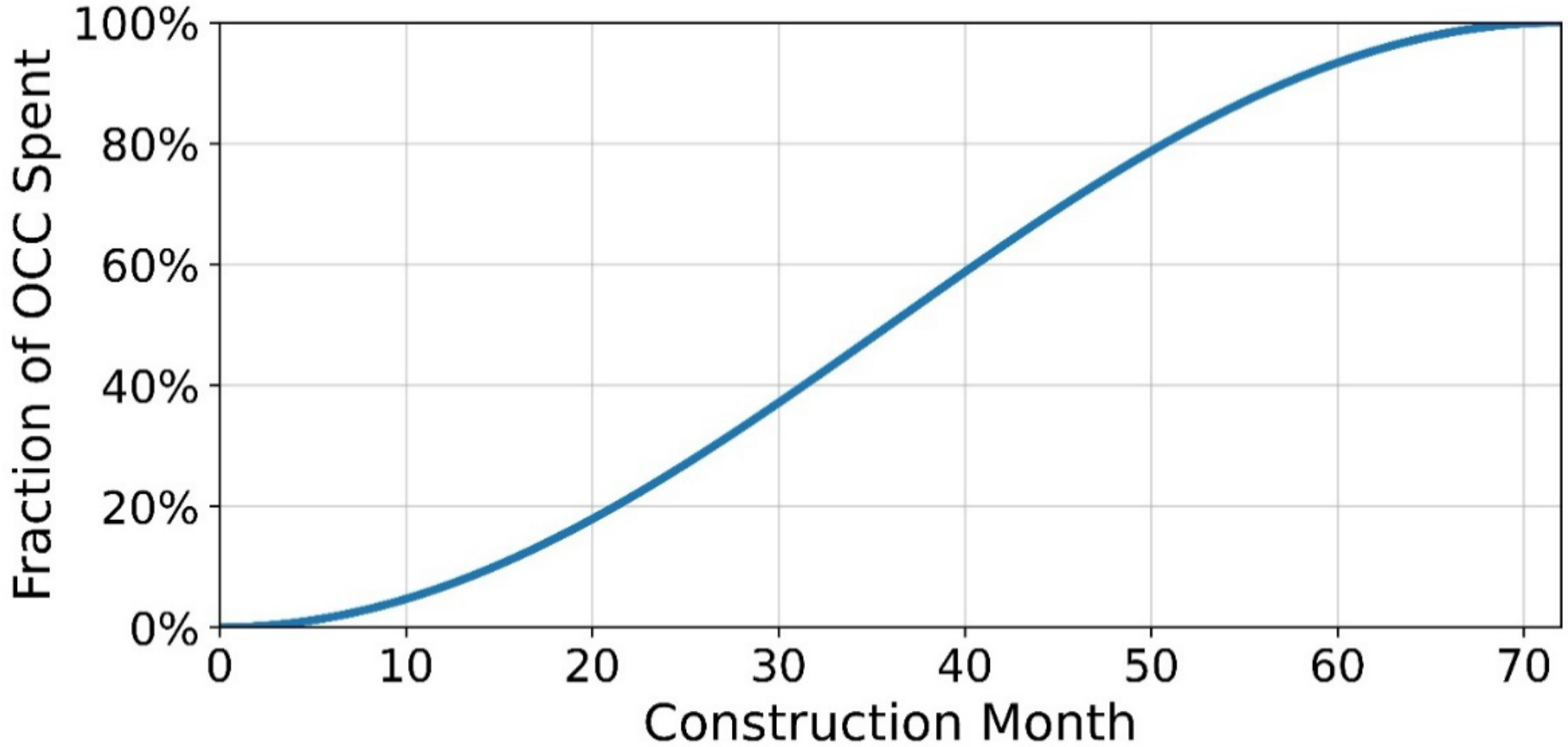

**Figure S1. Modeled cumulative OCC spending fraction for the PWR12-BE reactor cost model, with a construction duration of 72 months.**

The rate at which OCC is being spent, in 1987 USD per year, at any given time $t$ during construction can therefore be represented as:

$$\frac{dS}{dt} = OCC \times \frac{dF_t}{dt} = \frac{OCC}{2} \times \frac{\pi}{T} \times \sin\left(\frac{\pi t}{T}\right) \quad \text{(Equation S2)}$$

Thus, the interest during construction (IDC) accrued during an infinitesimal timestep $dt$ at time $t$ is given by:

$$IDC_t = r(T - t)\frac{dS}{dt}dt = r(T - t)dS \quad \text{(Equation S3)}$$

Where $r$ is the financing rate (4% per year for our modeling scenario – see *Results* § *Input assumptions*), $T - t$ is the number of years remaining until construction is complete, and $dS$ is the infinitesimal amount of OCC expended at time $t$. We then integrate Equation S3 on a continuous basis from $t$ = 0 to $t$ = $T$ to get the total financing cost as a function of the overnight capital cost and construction duration:

$$C_{fin}(OCC, T) = \int_0^T r(T-t)\frac{dS}{dt}dt = \frac{OCC}{2} \times \frac{\pi}{T} \times r \int_0^T (T-t)\sin\left(\frac{\pi t}{T}\right)dt \qquad \text{(Equation S4)}$$

Which finally yields:

$$C_{fin}(OCC, T) = \frac{OCC}{2} \times \frac{\pi}{T} \times r \times \frac{T^2}{\pi} = \frac{OCC}{2} \times T \times r \qquad \text{(Equation S5)}$$

The financing overrun ($\Delta C_{fin}$) between the PWR12-BE and PWR12-ME was therefore calculated as follows:

$$\Delta C_{fin} = C_{fin}(OCC, T) - C_{fin}(OCC_0, T_0) \qquad \text{(Equation S6)}$$

Where $C_{fin}(OCC, T)$ is the financing cost for the PWR12-ME ($OCC$ = \$2.529 billion (1987 USD), $T$ = 98 months) and $C_{fin}(OCC_0, T_0)$ is the financing cost for the PWR12-BE ($OCC_0$ = \$1.456 billion (1987 USD), $T$ = 72 months). Using the financing rate of $r$ = 4% we assume for our analysis (see *Results* § *Input assumptions*), we get $C_{fin}(OCC, T)$ = \$413 million and $C_{fin}(OCC_0, T_0)$ = \$175 million, so the financing overrun is $\Delta C_{fin}$ = \$238 million.

## Rework factor calculations

Equations 3 and 4 are derived from a sensitivity analysis which examines how the total capital investment (TCI) of the reactor cost model varies as a function of rework factor, a process described in Appendix C of the original Cost Reduction Tool report [14] that will also be described here for convenience. In the sensitivity analysis, Equations 3 and 4 are calibrated based on the assumption that an inexperienced A-E firm can increase TCI by 30%, so the TCI in the reactor cost model should increase by 30% when $AEP$ = 0, should increase by 0% when $AEP$ = 2, and $R_{AE}$ is assumed to vary linearly with $AEP$ between these two extremes. For the PWR12-BE with $p$ = 1.0 and $r$ = 4%, a TCI increase of 30% corresponds to a rework factor of 1.253 as reflected in Figure S2, hence the choice of coefficients in Equation 3. The Cost Reduction Tool also assumes the relation between $CSP$ and $R_C$ in Equation 4 is the same as that between $AEP$ and $R_{AE}$ [14]. Splitting these effects into two separate rework factors allows the user to make their own assumptions about how A-E proficiency and construction service proficiency each individually affect cost outcomes under different deployment scenarios.

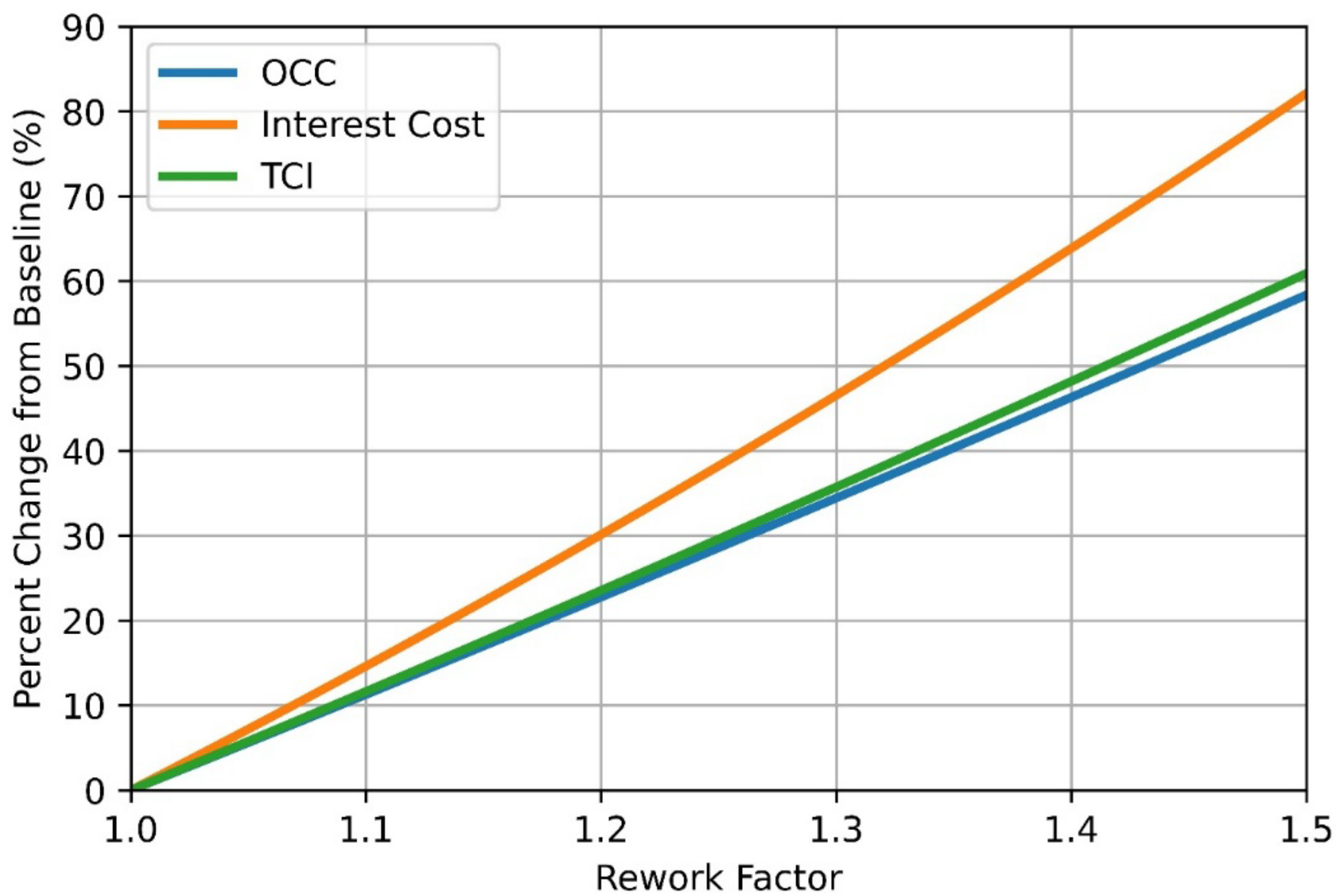

**Figure S2. Plot showing the sensitivity of PWR12-BE costs (OCC, financing, and TCI) to increases in rework factor.**

In the Cost Reduction Tool, the rework factor related to an incomplete design ($R_{design}$) is calculated as a function of a user-specified value for the level of design completion of the nuclear power plant prior to construction start, which may range from 0% to 100%. For the purpose of this analysis, we do not specify a value for design completion; instead, we calculate what the value of $R_{tot}$ must be to get from PWR12-BE costs to PWR12-ME costs after low productivity and supply chain delay overruns are accounted for (see the *Low Productivity* and *Supply Chain Delays* sections of *Methods*). Under the input assumptions used in our analysis, the resulting value is $R_{tot}$ = 1.327. The implied value of $R_{design}$ can then be back-calculated from $R_{tot}$, which results in $R_{design}$ = $R_{tot}$ / ($R_C \times R_{AE}$) = 1.107.

## Indirect cost scaling

Stewart and Shirvan proposed relations for how indirect costs scale with direct costs [1]:

$$C_{indirect,labor} = 0.36 \times C_{direct,labor} \quad \text{(Equation S7)}$$

$$C_{indirect,material} = 0.79 \times \frac{\text{average \# workers}}{\text{PWR12 BE average \# workers}} \times C_{direct,material} \quad \text{(Equation S8)}$$

$$C_{indirect,factory} = 1.32 \times \frac{\text{construction time}}{\text{PWR12 BE construction time}} \times C_{direct,labor} \quad \text{(Equation S9)}$$

Where “average # workers” is the average number of workers on the construction site for a whatever nuclear construction project is being analyzed, calculated as:

$$\text{average \# workers} = \frac{\text{direct labor hours}}{\text{construction months} \times 160} \quad \text{(Equation S10)}$$

The factor of 160 is the assumed average number of working hours for construction workers in a month. In general, direct and indirect labor hours are updated using the same method as direct and indirect labor costs (see Equation S7 and the last term in Equation 1) since these quantities can be assumed to be directly proportional to each other. In Equations S8 and S9, “PWR12-BE average # workers” and “PWR12-BE construction time” refer to the average number of workers on site for the PWR12-BE model (1,058, calculated using Equation S10) and the PWR12-BE construction time (72 months), respectively [8]. While Equations S7 - S9 are meant to be applicable to any reactor design, they are originally derived from the PWR12-BE and PWR12-ME cost models, making them the ideal choice for indirect cost scaling in our analysis.

## Labor productivity data

As noted in *Methods* § *Overruns by causer*, we used data from a 1980 survey of U.S. nuclear power plant construction workers to model each stakeholder’s assumed share of responsibility for overruns related to low labor productivity. We started by deciding which stakeholders could have caused each of the six categories of low productivity indicated in the survey (material availability, tool availability, crew interfacing, overcrowded work areas, instructions time, and inspection delays). The resulting matrix in Table S1 assumes that Design & Management could have been at least partially responsible for five of the six categories, and Construction Subcontractors for four of the six categories, so these two stakeholders should cause a disproportionately large portion of low productivity-related cost overruns.

**Table S1. Matrix used to attribute low productivity overruns to different stakeholders.** In the "Possible Causers" columns, a value of 1 indicates that the stakeholder may have been at least partially responsible for the corresponding source of lost productivity, while a value of 0 indicates that they could not have been responsible.

| | | | **Possible Causers** | | | |
|---|---|---|---|---|---|---|
| **Sources of Lost Productivity (Not Including Rework)** | Hours per Week | Percent of Non-Rework Unproductive Hours | Construction Subcontractors | Design & Management | Equipment Suppliers | Regulator |
| Material Availability | 6.80 | 28% | 1 | 1 | 1 | 0 |
| Tool Availability | 4.28 | 18% | 1 | 0 | 0 | 0 |
| Crew Interfacing | 3.54 | 15% | 1 | 1 | 0 | 0 |
| Overcrowded Work Areas | 4.62 | 19% | 0 | 1 | 0 | 0 |
| Instructions Time | 2.27 | 9% | 1 | 1 | 0 | 0 |
| Inspection Delays | 2.61 | 11% | 0 | 1 | 0 | 1 |
| TOTAL | 24.12 | 100% | 4 | 5 | 1 | 1 |

This information can be used to determine the minimum and maximum possible responsibility of each stakeholder for non-rework-related unproductive hours. The minimum possible responsibility is determined by assuming the stakeholder caused only the low productivity in the categories for which they are the only possible causer (e.g., Design & Management was assumed to be the only stakeholder who could be responsible for Overcrowded Work Areas, so at minimum, they caused the 4.62 hours per week of unproductive hours associated with this category). Maximum possible responsibility is determined by assuming the stakeholder caused 100% of the overruns in all categories for which they are a possible causer.

The results of this exercise are shown in Table S2. Since we do not know exactly where each stakeholder falls within their potential range of responsibility for low productivity, we simply assume the midpoint value for each stakeholder, as illustrated in Figure 6. Other sensitivities could be examined in future work. Note that the values of the four midpoints in Table S2 sum to 114%, so the midpoints were normalized by dividing each one by the sum of all midpoints, and the resulting normalized values were used to calculate shares of responsibility for low productivity.

**Table S2. Calculated shares of responsibility for productivity-related overruns among the four stakeholders involved in productivity losses.**

| | Construction Subcontractors | Design & Management | Equipment Suppliers | Regulator | TOTAL |
|---|---|---|---|---|---|
| Minimum Possible Responsibility for Non-Rework Unproductivity | 18% | 19% | 0% | 0% | N/A |
| Maximum Possible Responsibility for Non-Rework Unproductivity | 70% | 82% | 28% | 11% | N/A |
| Midpoint | 44% | 51% | 14% | 5% | 114% |
| Midpoint (Normalized to 100%) [a] | 38% | 44% | 12% | 5% | 100% |

[a] The midpoint for each stakeholder had to be normalized so that the total responsibility added up to 100 percent, as the original total was 114 percent.

It should be noted that the definition of unproductivity as listed in Table S1 and the definition of $p$ used in our calculations are somewhat different. Sebastian and Borcherding [15] examined hours per week spent on unproductive activities (i.e., activities other than the construction workers' craft), whereas in INL's Cost Reduction Tool, productivity is defined [14] as the inverse of the "Cost Performance Index" ($CPI$) used in the Vogtle Construction Monitoring report series [16] [17]:

$$p = \frac{1}{CPI} = \frac{hours\ earned}{hours\ spent} \quad \text{(Equation S11)}$$

Thus, $p$ takes values between 0 and 1, with 1 representing full achievement of a theoretical maximum productivity level for a well-executed project. Equation 9 is used to attribute productivity-related overruns ($\Delta C_{LP,i}$) even though the definition of unproductivity used to calculate $f_{LP,i}$ is different from the definition of unproductivity used to calculate $\Delta C_{LP,tot}$. This approach is acceptable because a follow-up study by Garner and Borcherding [S2] revealed that the breakdown in sources of unproductivity was quite consistent across the 11 nuclear power plants surveyed (see Figure S3), even though the level of absolute unproductivity varied widely between the 11 plants in the dataset (Figure S4).

The project completion level (ranging from ~4% to ~77%) and locations of the plants were also very different and yet the breakdowns in sources of unproductivity remained similar across all projects. This gave us additional confidence that we can attribute responsibility for productivity-related overruns using constant $f_{LP,i}$ values regardless of the value of $\Delta C_{LP,tot}$ or the productivity input we assume.

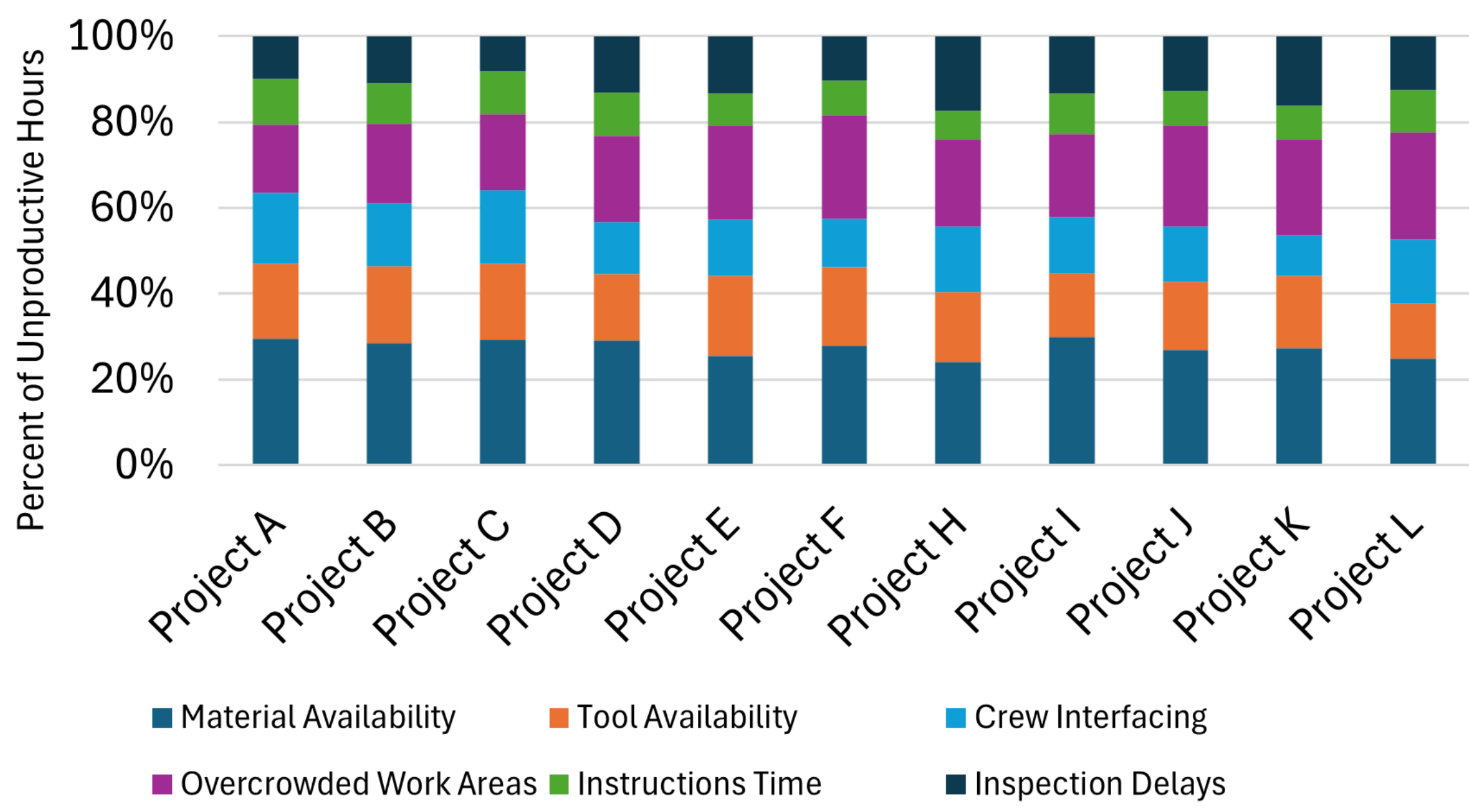

**Figure S3. Sources of non-rework unproductive hours across the 11 U.S. nuclear construction projects surveyed by Borcherding et al. [S2].**

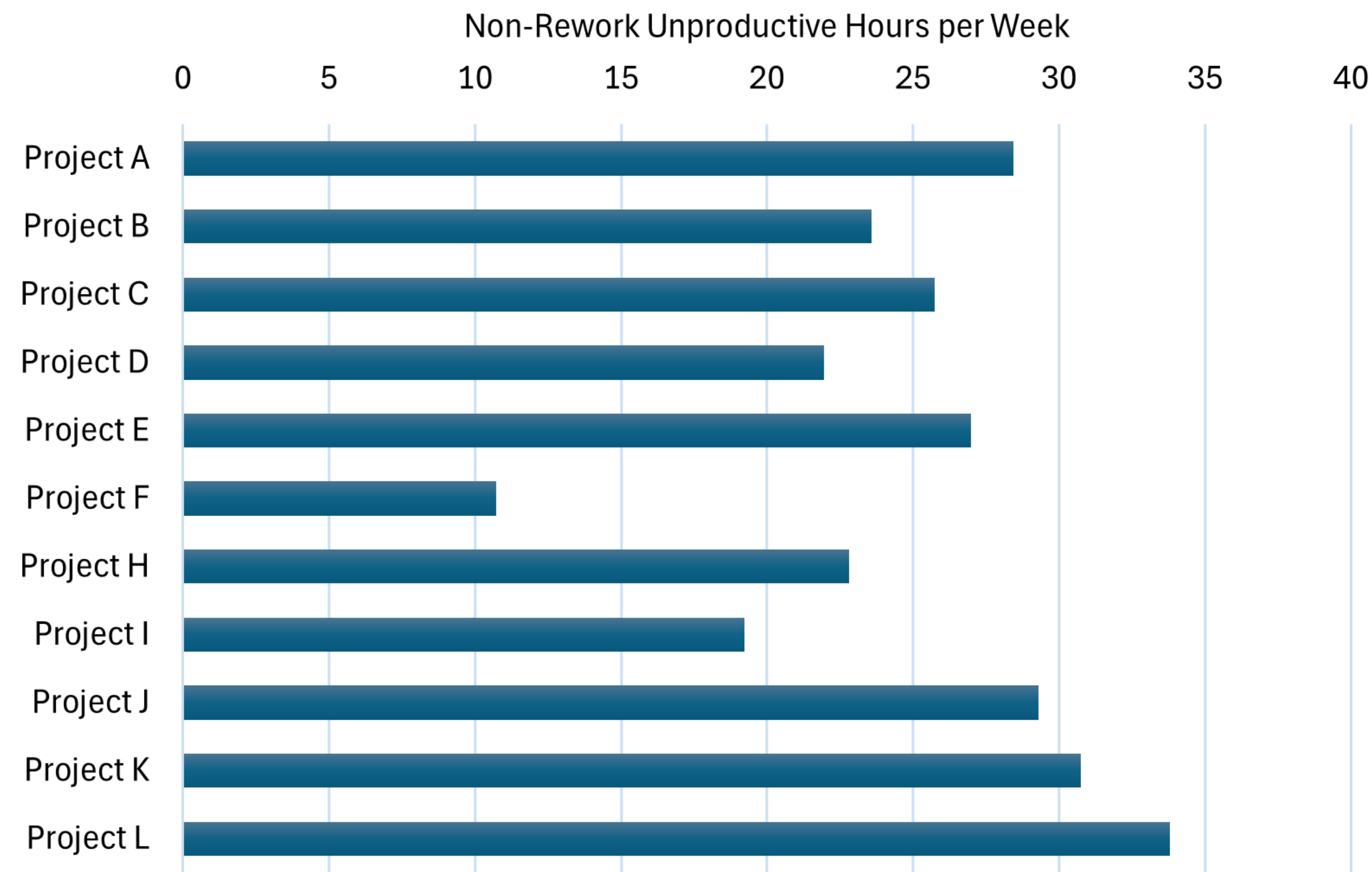

**Figure S4. Total non-rework unproductive hours per 40-hour work week across the 11 U.S. nuclear construction projects surveyed by Borcherding et al. [S2].**

## Profit vs. overrun curves

In *Results*, a "profit vs. overrun curve" is generated from the proposed performance-based contract framework shown in Figure 5(A). It may seem surprising that the resulting profit vs. overrun curve in Figure 5(B) is quadratic when the contract curve used to generate it is linear. The quadratic shape arises because costs and profits are defined differently for the owner than they are for the stakeholders directly involved in delivering the plant: payments from the owner to another project stakeholder represent only costs for the owner, but include both costs and profits for the stakeholder. Thus, in calculating the profit vs. overrun curve, the total profit received by the stakeholder is calculated as:

$$profit(OR) = pm(OR) \times [WE + OR] \quad \text{(Equation S12)}$$

Where $pm$ is the profit margin applicable to the stakeholder, $WE$ is the "well-executed" (or "baseline") project cost charged by the stakeholder in the case of the PWR12-BE (see *Determination of baseline cost scopes*), and $OR$ is the cost overrun in the stakeholder's cost scope. As Equation S12 indicates, $pm$ is a function of $OR$. Since the $pm(OR)$ function used to generate the profit vs. overrun curve in Figure 5(B) is the linear function shown in Figure 5(A), and this first-order $OR$ term is multiplied by $OR$ in Equation S12, the result is a curve that is second-order with respect to $OR$.

By contrast, the profit vs. overrun curves for the fixed-price and cost-plus contract cases are both linear. As mentioned in *Results*, the default assumption for the fixed-price contract is a 30% contingency for overruns, and the profit margin is set at 8% for this level of overruns. Thus, the "baseline profit" at 30% overruns is given by Equation S13:

$$baseline\ profit = 0.08 \times (WE \times 1.3) \quad \text{(Equation S13)}$$

Since costs charged to the project owner include profits for the stakeholders they are paying, Equation S13 implicitly assumes that at a 30% level of cost overruns, 8% of the project costs calculated within the Cost Reduction Tool consist of profits for the stakeholders, while the remaining 92% is internalized costs. Unfortunately, the EEDB code of accounts does not break down costs into internalized costs and profits, so it is difficult to determine how close this assumption is to reality. Nonetheless, the assumption is used for the sake of this illustrative example, and 8% does align with the average of the net profit margins for the industries of Machinery (10.04%), Engineering / Construction (2.95%), and Construction Supplies (11.23%) [S3].

The actual profit received by the stakeholder is then determined by the level of excess overruns over the expected contingency of 30%:

$$profit(OR) = baseline\ profit - excess\ overruns = baseline\ profit - (OR - 0.3 \times WE) \quad \text{(Equation S14)}$$

The result is a profit vs. overrun curve that is linear with respect to $OR$. Plugging overrun values into Equation S14, it can be seen that the expected profit at 0% cost overruns is $0.404 \times WE$ (a 40.4% profit margin), and at 60% overruns is $-0.196 \times WE$ (a profit margin of $-0.196 \times WE/(WE + OR) = -0.196 \times WE/(1.6 \times WE) = -0.196/1.6 = -12.25\%$), as reflected in Table 2.

Finally, the equation for the profit margin in the cost-plus contract case is similar to Equation S12, but with $pm(OR)$ held fixed as a function of $OR$ at a constant value of 8 percent:

$$profit(OR) = pm \times [WE + OR] = 0.08 \times [WE + OR] \quad \text{(Equation S15)}$$

Hence, the profit vs. overrun curve is first-order with respect to overruns and appears linear in Figure 5.

## Determination of baseline cost scopes

The "baseline cost scope" of each stakeholder is the amount of payments they receive in the baseline case of the PWR12-BE reactor cost model with zero overruns. We use the same methodology as described in *Methods* § *Overruns by recipient* to split the PWR12-BE direct costs between the Equipment Suppliers and Construction Subcontractors, giving them \$518 million and \$387 million respectively (1987 USD). For indirect costs, since 41.2% of the PWR12-BE indirect costs are in Account 91 (Construction Services), we assign these costs to the Construction Subcontractors, and the remaining 58.8% in Account 92 (Engineering and Home Office Services) and Account 93 (Field Supervision and Field Office Services) to Design & Management, which gives these stakeholders \$227 million and \$325 million respectively. Thus, the total baseline cost scopes for each stakeholder are \$518 million for Equipment Suppliers, \$614 million for Construction Subcontractors, and \$325 million for Design & Management.

## Construction of waterfall plots

To construct the waterfall plots in Figure 2, we needed to determine the individual impacts of rework, low productivity, and supply chain delays on OCC and construction duration. Since Equations 1 - 5 are interdependent, it is difficult to immediately determine the individual shares of OCC and schedule overruns caused by the input values of $R_{tot}$ (for rework effects), $p$ (for low productivity effects), and the supply chain delay, $\Delta T_{SCD}$. To address this, we used a 3-factor Shapley decomposition to determine the shares of overruns caused by each of these three effects:

$$\phi_P = \frac{V_P - V_0}{3} + \frac{V_{PR} - V_R}{6} + \frac{V_{PS} - V_S}{6} + \frac{V_{PRS} - V_{RS}}{3} \quad \text{(Equation S16)}$$

$$\phi_R = \frac{V_R - V_0}{3} + \frac{V_{PR} - V_P}{6} + \frac{V_{RS} - V_S}{6} + \frac{V_{PRS} - V_{PS}}{3} \quad \text{(Equation S17)}$$

$$\phi_S = \frac{V_S - V_0}{3} + \frac{V_{PS} - V_P}{6} + \frac{V_{RS} - V_R}{6} + \frac{V_{PRS} - V_{PR}}{3} \quad \text{(Equation S18)}$$

Where $V$ represents either the OCC or construction duration calculated by the stakeholder analysis framework (Equations 1 - 5) under a certain set of inputs, $\phi$ is the share of OCC or construction duration attributed to each of the three overrun mechanisms, and the subscript letters $P$ (productivity), $R$ (rework), and $S$ (supply chain) are used to represent what combination of inputs is used in each case. A subscript "$P$" indicates that the input value for productivity is set to $p$ = 0.75, the default for our analysis, while $R_{tot}$ = 1 and $\Delta T_{SCD}$ = 0, indicating no effects from rework or supply chain delays. A subscript "$PR$" indicates $p$ = 0.75, $R_{tot}$ = 1.327 (the default for the analysis), and $\Delta T_{SCD}$ = 0 (no effect). A subscript "$PRS$" indicates $p$ = 0.75, $R_{tot}$ = 1.327, and $\Delta T_{SCD}$ = 9.4 months. All other subscript combinations are defined similarly, and the subscript "$0$" indicates that none of the three overrun effects are applied, so the baseline value of OCC or construction duration from the PWR12-BE is used.

## Supplemental References